\documentclass[sigconf, nonacm]{acmart}
\usepackage{enumitem}
\usepackage{multirow}
\usepackage{graphicx}
\usepackage{dirtytalk}
\usepackage{subcaption}
\usepackage{graphicx}
\AtBeginDocument{%
  \providecommand\BibTeX{{%
    \normalfont B\kern-0.5em{\scshape i\kern-0.25em b}\kern-0.8em\TeX}}}

\setcopyright{acmcopyright}
\copyrightyear{2024}
\acmYear{2024}
\acmDOI{XXXXXXX.XXXXXXX}

\acmConference[]{}

\acmSubmissionID{}
 
\usepackage{booktabs}
\usepackage{multirow}
\usepackage[utf8]{inputenc}
\usepackage{array}
\usepackage[table]{xcolor}
\usepackage[most]{tcolorbox}
\usepackage{caption}
\usepackage{xspace}

\newcommand{\touchOne}{1H-thumb\xspace}
\newcommand{\touchTwo}{2H-index\xspace}
\newcommand{\gazeF}{Gaze-F\xspace}
\newcommand{\gazetouchF}{GazeTouch-F\xspace}

\begin{document}

\settopmatter{printccs=false}

\definecolor{red1}{RGB}{255, 230, 230} % Lightest red
\definecolor{red2}{RGB}{255, 204, 204}
\definecolor{red3}{RGB}{255, 153, 153}
\definecolor{red4}{RGB}{255, 102, 102}
\definecolor{red5}{RGB}{255, 51, 51}
\definecolor{red6}{RGB}{220, 40, 40}
\definecolor{lightturquoise}{RGB}{173, 216, 230} % Light turquoise
\definecolor{darkerturquoise}{RGB}{123, 176, 190}
\definecolor{shadecolor}{RGB}{255, 204, 204}
\definecolor{lightgreen}{rgb}{0.9, 1.0, 0.9} % Light green

\newcounter{observation}
\renewcommand{\theobservation}{\arabic{observation}}

% the box style
\newtcolorbox{observationbox}{
  enhanced, breakable,
  colback=red2, colframe=red2,
  boxrule=0.8pt, arc=1mm,
  left=5pt, right=5pt, top=5pt, bottom=5pt,
  before skip=\baselineskip, after skip=\baselineskip
}

\newcommand{\observation}[1]{%
  \refstepcounter{observation}%
  \begin{observationbox}
    \textbf{Observation~\theobservation: }#1
  \end{observationbox}%
}

%%
%% The "title" command has an optional parameter,
%% allowing the author to define a "short title" to be used in page headers.

% \title [the Effect of Encumbrance on Target Acquisition Using Gaze]{Investigating the Effect of Encumbrance on Target Acquisition Using Gaze on Handheld Mobile Devices}

\title [Investigating the Effect of Encumbrance on Gaze- and Touch-based Target Acquisition]{Investigating the Effect of Encumbrance on Gaze- and Touch-based Target Acquisition on Handheld Mobile Devices}

%%
%% The "author" command and its associated commands are used to define
%% the authors and their affiliations.
%% Of note is the shared affiliation of the first two authors, and the
%% "authornote" and "authornotemark" commands
%% used to denote shared contribution to the research.
\author{Omar Namnakani}
\email{o.namnakani.1@research.gla.ac.uk}
\orcid{0000-0002-3803-5781}
\affiliation{%
  \institution{University of Glasgow}
  \country{United Kingdom}
}

\author{Yasmeen Abdrabou}
\email{yasmeen.essam@unibw.de}
\orcid{0000-0002-8895-4997}
\affiliation{%
  \institution{Human-Centered Technologies for Learning,\\ Technical University of Munich}
  \city{Munich}
  \country{Germany}
}

\author{John H. Williamson}
\orcid{0000-0001-8085-7853}
 \email{johnh.williamson@glasgow.ac.uk}
\affiliation{%
\institution{University of Glasgow}
  \country{United Kingdom}
} 

\author{Mohamed Khamis}
\email{mohamed.khamis@glasgow.ac.uk}
\orcid{0000-0001-7051-5200}
\affiliation{%
  \institution{University of Glasgow}
  \country{Glasgow, United Kingdom}
}  

%%
%% By default, the full list of authors will be used in the page
%% headers. Often, this list is too long, and will overlap
%% other information printed in the page headers. This command allows
%% the author to define a more concise list
%% of authors' names for this purpose.
\renewcommand{\shortauthors}{Namnakani et al.}

%%
%% The abstract is a short summary of the work to be presented in the
%% article.
\begin{abstract}
    The potential of using gaze as an input modality in the mobile context is growing. While users often encumber themselves by carrying objects and using mobile devices while walking, the impact of encumbrance on gaze input performance remains unexplored. To investigate this, we conducted a user study ($N=24$) to evaluate the effect of encumbrance on the performance of 1) Gaze using Dwell time (with/without visual feedback), 2) GazeTouch (with/without visual feedback), and 3) One- or two-hand touch input. While Touch generally performed better, Gaze, especially with feedback, showed a consistent performance regardless of whether participants were encumbered or unencumbered.
Participants' preferences for input modalities varied with encumbrance: they preferred Gaze when encumbered, and touch when unencumbered. Our findings enhance understanding of the effect of encumbrance on gaze input and contribute towards selecting appropriate input modalities in future mobile user interfaces to account for situational impairments.

\end{abstract}

%%
%% The code below is generated by the tool at http://dl.acm.org/ccs.cfm.
%% Please copy and paste the code instead of the example below.
%%

%%
%% Keywords. The author(s) should pick words that accurately describe
%% the work being presented. Separate the keywords with commas.
\keywords{Gaze-based Interactions, Mobile Devices; Dwell time; Touch input; Multimodal Interaction, gaze-supported interaction, Encumbrance}

%% A "teaser" image appears between the author and affiliation
%% information and the body of the document, and typically spans the
%% page.

% \received{12 September 2024}
% \received[revised]{12 March 2009}
% \received[accepted]{5 June 2009}

\begin{teaserfigure}
  \includegraphics[width=\textwidth]{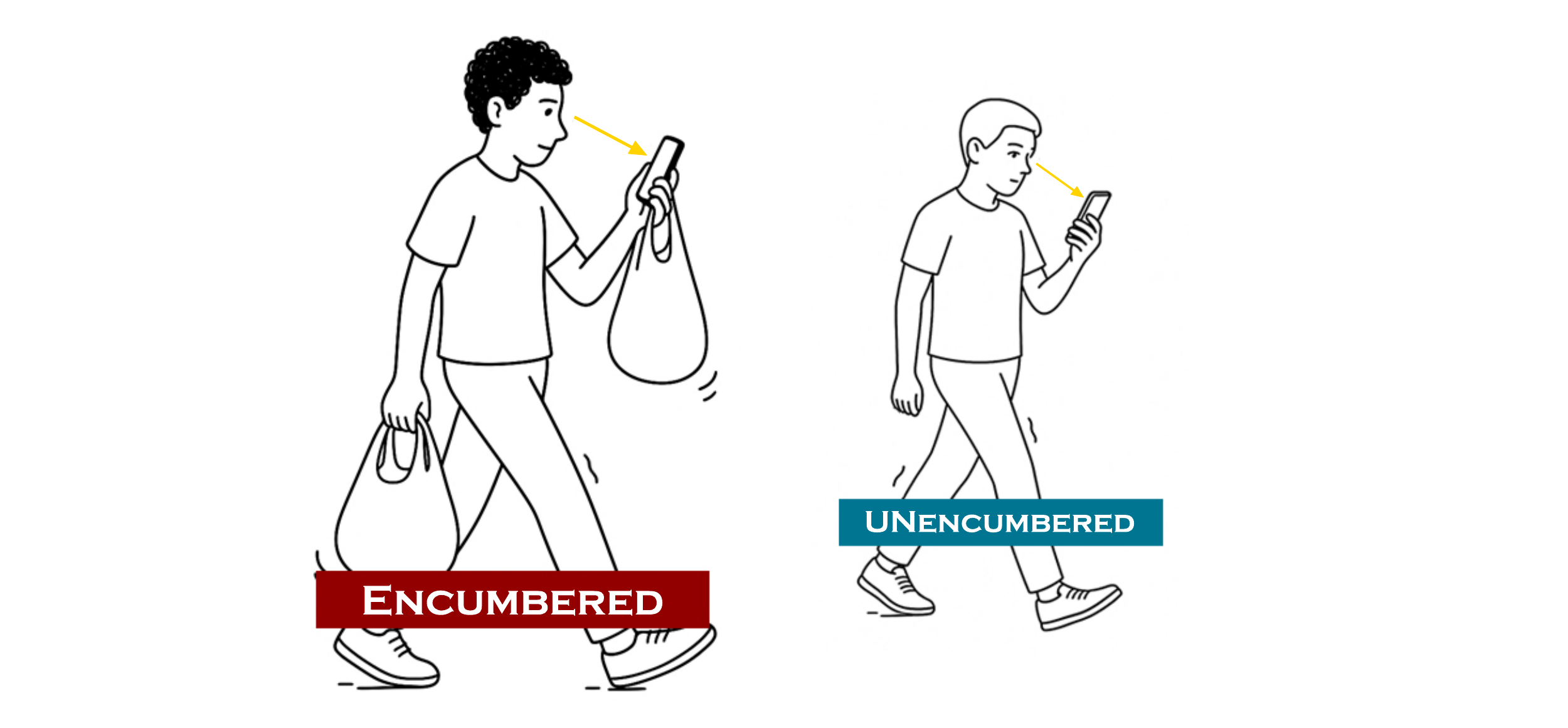}
   \caption{We evaluate gaze- and touch-based target selection on handheld mobile devices while walking encumbered (carrying bags) and unencumbered. Besides using gaze alone, we examine multimodal interaction combining gaze for pointing and touch for selection, with touch input included as a baseline. Left: a participant walking while carrying two bags and interacting using gaze (Encumbered). Right: a participant walking and interacting using gaze without carrying bags (Unencumbered).}
  \label{fig:teaser}
  \Description{}
\end{teaserfigure}

%%
%% This command processes the author and affiliation and title
%% information and builds the first part of the formatted document.
\maketitle

\section{Introduction}
Gaze interaction on mobile devices is an emerging field of research that investigates the use of eye movements to operate smartphones and tablets~\cite{10.1145/3229434.3229452}. It holds promise for becoming a natural part of everyday interaction~\cite{Møllenbach_Hansen_Lillholm_2013, 10.5555/1778331.1778385}
Given the current advances in front-facing cameras and processing power of the said devices, eye tracking on handheld mobile devices is advancing \cite{10.1145/3229434.3229452}, with recent work exploring the use of gaze as an input modality on handheld mobile devices by evaluating gaze interaction techniques in the mobile context \cite{10.1145/3706598.3713092, 10.1145/3544548.3580871, lei2023dynamicread}. Using gaze as an input modality can be advantageous, particularly in situations where touch interaction is limited, such as while walking, holding objects, while wearing gloves, or when the screen is wet.~\cite{10.1145/3152771.3156161, 10.1145/1409240.1409253, goncalves2017tapping}.

Prior work showed that users move through the environment and carry objects daily, which encumbers them while walking and interacting with their mobile devices \cite{Ng2013Encumbrance}. Although touch is the mainstream input modality for mobile devices \cite{Orphanides2017Touchscreen}, walking while encumbered has been shown to negatively affect its performance, as accuracy decreases~\cite{10.1145/2556288.2557312, 10.1145/2785830.2785853}. Given that prior work suggested using alternative modalities depending on the mobile context~\cite{saulynas2022putting}, advancing gaze as an alternative input when users are encumbered requires exploring the effect of encumbrance on its performance, as prior work suggested that new input modalities should be tested for encumbrance~\cite{ng2011effects}. 
In this paper, we investigate the effects of encumbrance on Gaze input using dwell time, which is the most common gaze interaction technique evaluated in the literature \cite{10.1145/123078.128728, Møllenbach_Hansen_Lillholm_2013, 10.1145/1378063.1378122, 10.1145/97243.97246}. Dwell time was introduced to mitigate the Midas touch problem, and showed promising results in the mobile context, where users preferred it \cite{lei2023dynamicread, 10.1145/3544548.3580871} compared to other techniques, such as Pursuits \cite{VidalPusuits2013, 10.1145/2807442.2807499} and gestures \cite{10.1145/1743666.1743710, 10.1145/2556288.2557040}. 
Besides using Dwell time, we also explore the impact of encumbrance on multimodal interaction, when using gaze in combination with touch (GazeTouch), as prior work suggested that gaze has the potential to complement other modalities \cite{10.1145/2642918.2647397}, where gaze is used for pointing and touch for selection \cite{10.1145/2642918.2647397, Pfeuffer2015GazeTouch,10.1145/2858036.2858201}. 
Additionally, we incorporated touch input using two common postures as baselines: \touchOne (one-handed use, where the user holds the phone using their preferred hand and interacts with the thumb), and \touchTwo (two-handed use, where one hand holds the phone and the other interacts with the index finger), enabling comparisons of users’ perceptions and preferences under different levels of encumbrance.

In a user study with $24$ participants, for a discrete selection task on a handheld mobile device, we evaluated the impact of encumbrance on the performance of three input modalities when participants performed selection while walking encumbered and walking unencumbered, in terms of selection time, task completion rate, error counts, perceived workload (NASA-TLX), perceived usability (Likert scale), and user preference. Participants used three input modalities 1) Gaze input using dwell time with visual feedback (\gazeF) and without (Gaze), 2) touch when combined with gaze, where gaze is used for pointing and touch for manipulating, with visual feedback (\gazetouchF) and without (GazeTouch), and 3) Touch input, one handed (\touchOne) and two handed (\touchTwo).

Based on statistical tests using ANOVA and Bayesian analysis, we found that the performance of both \touchOne and \touchTwo was affected by encumbrance, where participants took a longer time to select, completed fewer tasks, and committed more errors when walking encumbered compared to when unencumbered.
On the other hand, Gaze, especially with feedback (\gazeF), showed comparable performance between walking encumbered and unencumbered. GazeTouch with and without feedback showed comparable selection time and error counts, but participants completed less tasks when encumbered compared to when unencumbered. 
Participants perceived \gazetouchF, \touchOne, and \touchTwo to cause significantly higher workload when encumbered than when unencumbered.
Participants' preferences for input modalities varied with encumbrance: they preferred Gaze when encumbered, and touch when unencumbered. 
Our findings enhance our understanding of the effect of encumbrance on gaze input and contribute towards selecting appropriate interaction techniques to counteract the problem of such encumbrance when interacting with handheld mobile devices.

\subsection*{Contribution Statement}
This work provides two contributions. We quantify, for the first time, the effects of encumbrance and walking on gaze-based target selection for handheld mobile devices. 
Second, we present the first empirical results, showing that gaze may have the potential to be used for target selection on handheld mobile devices, especially when touch-based interaction is limited, highlighting the importance of not excluding gaze solely based on objective performance.

\section{Background and Related Work}
Our work builds on prior research on gaze interaction on handheld mobile devices, touch-based interaction, multimodal gaze-based interaction, and on situational impairments induced by the mobile context.

\subsection{Encumbrance, Mobility, and Situationally Induced Impairments to Touch Input on Handheld Mobile Devices}
Prior work has explored the impact of encumbrance, such as when users carry objects, a frequent situation in everyday life, on the performance of interaction techniques, for example, wrist rotation gestures \cite{ng2011effects}. The researchers suggested that designers should test their interaction techniques with a range of different encumbrances to ensure that they are useful in real-world contexts. 
Following that and given the popularity of touch screen mobile devices, a plethora of other work focused on the performance of touch input on smartwatches and mobile devices when users are encumbered \cite{10.1145/2556288.2557312, 10.1145/2628363.2628382, 10.1145/3123021.3123033, 10.1145/3173574.3174208}. Encumbrance by carrying bags while walking was observed by Ng et al.~\cite{Ng2013Encumbrance} as frequently occurring in public during interaction with mobile devices. Based on their observational study, they conducted a user study to measure the performance of touch input for target selection on mobile devices when users are walking encumbered \cite{10.1145/2556288.2557312}. They found that the accuracy for selection significantly dropped when users were encumbered compared to when holding no objects while on the move, and that the encumbrance affected the dominant hand more than the non-dominant one. Later, Ng et al.~ examined two walking evaluation methods, either on the ground or on a treadmill, to better investigate the effects of encumbrance while the preferred walking speed (PWS) is controlled \cite{10.1145/2628363.2628382}. They found that walking on the ground yielded a better representation of participants' PWS than did treadmill walking. Following that, the researchers examined the effects of mobility (walking) and encumbrance on standard touchscreen gestures, including tapping, dragging, spreading \& pinching, and rotating clockwise \& anticlockwise \cite{10.1145/2785830.2785853}. Their results showed that tapping and dragging performed poorly when users were encumbered and on the move. They suggested that designers should think about what input techniques to use in everyday mobile settings, where users are more likely to be carrying bags. 
Beside encumbrance, prior work has shown that touch input performance can be impacted in various other situations, such as when users are walking \cite{10.1145/1409240.1409253, 10.1145/3319499.3330292}, wearing impeding clothes such as gloves \cite{10.1145/3319499.3330292}, the screen is wet \cite{10.1145/3242969.3243028}, when using handheld mobile devices in cold temperatures \cite{goncalves2017tapping, 10.1145/2971648.2971734}, and depending on the users context \cite{10.1145/3577013}. Such impact on performance may lead to what is known as situationally induced impairments~\cite{wobbrock2019situationally, 10.1145/1409240.1409253}. The term refers to contextual factors that can negatively impact users of various abilities when interacting with mobile devices \cite{wobbrock2006future, 10.1145/3152771.3156161,wobbrock2019situationally, 10.1145/3386370}. The researchers noted that users experiencing situational impairments might interact with their touch devices similarly to how individuals with physical or sensory impairments do.

Given the preliminary design guidelines from prior work, which suggested using alternative input modalities that are appropriate for each mobile context~\cite{saulynas2022putting}, and the suggestion of gaze as a crucial input modality in scenarios of situational impairments \cite{10.1145/3204493.3208344}, in this work, we investigate the impact of encumbrance on gaze input using dwell time. Such an investigation will help us explore and uncover the potential of using gaze as an alternative modality to mitigate the impact of encumbrance and mobility during interaction with mobile devices, as these devices are used in a diverse range of contexts \cite{10.1145/2628363.2628382}.

\subsection{Gaze and Touch Interaction on Handheld Mobile Devices}
While touch input has been the mainstream input modality on handheld mobile devices \cite{Orphanides2017Touchscreen}, the demonstrated feasibility of gaze as an input modality by Jacob et al.~\cite{10.1145/332040.332445}, along with recent advances in mobile device processing power and front-facing cameras \cite{10.1145/3229434.3229452}, has begun to interest researchers in exploring the potential of using gaze on mobile devices \cite{10.1145/3544548.3580871, 10.1145/3706598.3713092, lei2023dynamicread, 10.1145/3591133}. Apple has introduced gaze using dwell time as an optional input modality in their recent iPhones \cite{apple_tracking}.

Dwell time for gaze input is the most common technique explored in the literature \cite{10.1145/2414536.2414609, 10.1145/332040.332445, 10.1145/97243.97246, 10.1145/123078.128728}. The technique introduces a brief time to fixate on the targets for selection, thereby mitigating the Midas touch problem by differentiating between the user's casual viewing and gaze input.
As an alternative to dwell time, Pursuits \cite{10.1145/2493432.2493477, VidalPusuits2013, 10.1145/2807442.2807499} and Gesuters \cite{10.1145/2168556.2168601, 10.1145/2168556.2168579, 10.1145/1743666.1743710, 10.5555/1778331.1778385} were introduced. Both techniques rely on relative eye movements rather than the absolute point of gaze, suggesting them as promising for the context of handheld mobile devices \cite{10.1145/3229434.3229452, esteves2020comparing}.
While these techniques were evaluated in settings that are different from mobile, such as desktop \cite{10.1145/2414536.2414609, 10.1145/123078.128728} and HMDs \cite{10.1145/3206505.3206522, esteves2020comparing}, recent work by Namnakani et al.~\cite {10.1145/3544548.3580871} compared dwell time, Pursuits, and Gestures for target selection in mobile settings, when participants were seated and on the move, with a varying number of targets. While they found that Pursuits was the fastest overall, the performance and preference for input techniques varied depending on the user's context and number of targets. Overall, their results showed that participants preferred Dwell time when walking and as the number of targets on the screen increased. Similarly, Lei et al.~\cite{lei2023dynamicread} explored gaze input for scrolling in the mobile context, demonstrating that dwell time was easier to control and more robust than Pursuits while walking.  
The preference for dwell time in the mobile context could be due to its ease of learning, its intuitiveness, and its role as the gaze counterpart to touch tap \cite{10.1145/3706598.3713092}.
As dwell time showed promising results in the mobile context, in this work, we aim to explore the impact of encumbrance on gaze interfaces that utilise dwell time for selection on handheld mobile devices.

Given that unintentional target selection remains a challenge in gaze-based interaction \cite{10.1145/2207676.2208709}, beyond using dwell time, researchers have explored combining gaze with another input modality to mitigate the Midas Touch problem. The second input modality could be voice \cite{10.1145/3490099.3511103}, hand gestures \cite{Gaze+pinch}, motion gestures \cite{kong2021eyemu}, keyboard and/or mouse \cite{salvucci2000intelligent, 10.1145/1054972.1054994}, or touch \cite{10.1145/2642918.2647397}. Combining gaze with touch has been explored extensively on interactive surfaces and when integrated into the same user interface, such as on tablets \cite{Pfeuffer2015GazeTouch,10.1145/2984511.2984514, 10.1145/2858036.2858201, 10.1145/2642918.2647397, 10.1145/2207676.2208709}. 
While prior work on large displays showed that users can perform comparably whether using touch or gaze \& touch \cite{Pfeuffer2015GazeTouch, 10.1145/2858036.2858201}, and that the technique allows for unimanual, single-grip, thumb-only interaction with tablets \cite{10.1145/2984511.2984514}, this motivated us also to explore the gaze \& touch technique in the mobile context, and examine the impact of encumbrance on its performance. The technique is based on the principle that \textit{gaze points, and touch manipulates}, where users can look at the target and touch anywhere on the screen to confirm the selection \cite{10.1145/2984511.2984514, 10.1145/2642918.2647397}.

\subsection{Gap Summary and Research Questions}
While prior work has explored the impact of encumbrance on touch input, the impact on gaze input, whether gaze is used alone or supported by touch in multimodal interaction, remains to be explored. Therefore, in this work, we focus on investigating the impact of encumbrance on target selection using dwell input for gaze-enabled interfaces on handheld mobile devices. We aim to answer the following research questions: 
\begin{enumerate}[label=\textbf{RQ\arabic*}, leftmargin=1.5cm]
  \item How does encumbrance impact the performance of gaze input techniques, compared to touch input? 
  \item How do participants perceive using gaze as an alternative input modality to touch while encumbered?
\end{enumerate}

\section{Concepts and Implementation}
In this section, we explain the concepts alongside our implementation of two gaze-based techniques used in our study: 1) Dwell time, and 2) the multimodal interaction, gaze \& touch (GazeTouch), where gaze is used for pointing and touch for selection. Touch input utilises the basic tap gestures found in Apple's iPhone devices, which are quick touches down and lifts of a single finger, recognised with a single tap for selection. 

\subsection{Gaze Using Dwell Time}
We enable gaze input using Dwell time. Dwell time requires users to fixate on a target for a period of time to perform a selection \cite{10.1145/97243.97246, 10.1145/123078.128728, 10.1145/3419249.3420122, 10.1145/3544548.3580871, Møllenbach_Hansen_Lillholm_2013}.
In our implementation of Dwell time, we used a dwell duration of $500\,ms$ based on prior work~\cite{ AbdrabouJustGaze2019, lei2023dynamicread, 10.1145/3419249.3420122, Hansen2003DwellTyping}. We confirmed that our decision was reasonable to fairly match the average time touch takes, as prior work showed that touch input takes approximately between $500-650\,ms$ for selection while walking \cite{10.1145/2785830.2785853}. Rather than requiring every gaze point to fall inside a target during this time window, we account for the dynamic nature of the mobile context, which introduces noise into gaze data. Specifically, once a gaze point enters a candidate target, it is considered a candidate target, provided at least 70\% of the gaze points remain within the target's boundary \cite{10.1145/3706598.3713092}. If this criterion is not met, the dwell timer is cancelled and the collection of gaze points is reset. If the criterion is met and the dwell timer fired, we verify whether the average of the collected gaze points within the time window lies inside the candidate target. If so, the target is confirmed as selected. This implementation allowed us to provide visual feedback (see Section \ref{sec:visualF}) to any candidate target that participants were looking at, while accounting for the inaccuracy of gaze tracking in mobile contexts.

\subsection{GazeTouch}
Prior work in the last decades explored GazeTouch, including \cite{10.1145/2207676.2208709} and \cite{10.1145/2642918.2647397}, based on the principle that gaze selects and touch manipulates and confirms. This multimodal interaction method was introduced to overcome the Midas touch problem while maintaining task performance \cite{10.1145/2207676.2208709}, as the method does not require a dwell time, which is time-consuming and slows down the interaction \cite{10.1145/1983302.1983303}. 
In our implementation of GazeTouch, the target to be selected is determined by the user's current point of gaze. When a user touches the screen by tapping anywhere, the target they are currently looking at is considered the candidate target, corresponding to the touch input. The selection is then confirmed and finalised the moment the user terminates the touch gesture by lifting their finger from the screen.   

\subsection{Visual Feedback for Gaze and GazeTouch} \label{sec:visualF}
While we provided discrete visual feedback for all input modalities once targets were selected, we added two additional variants of Gaze and GazeTouch that highlight targets currently in focus, because when using gaze, users depend on the system's interpretation of where they are looking (see Figure \ref{fig:screenshot}). Touch inherently includes such feedback, as users physically point and control exactly what they are selecting.
Prior work showed that visual feedback helps users confirm or cancel their selection before the dwell completes and is also reported by participants to give confidence in their selection, knowing which target was about to be selected \cite{10.1145/3419249.3420122, Majaranta2006, 10.1145/968363.968390}. We added visual feedback for Gaze by highlighting the target currently in focus with a red outline before it is selected \cite{10.1145/968363.968390}. For GazeTouch, participants received immediate feedback through a red outline highlighting the target they were looking at.  
\section{Evaluation}
\begin{figure*}[!t]
  \centering
  \includegraphics[width=\linewidth]{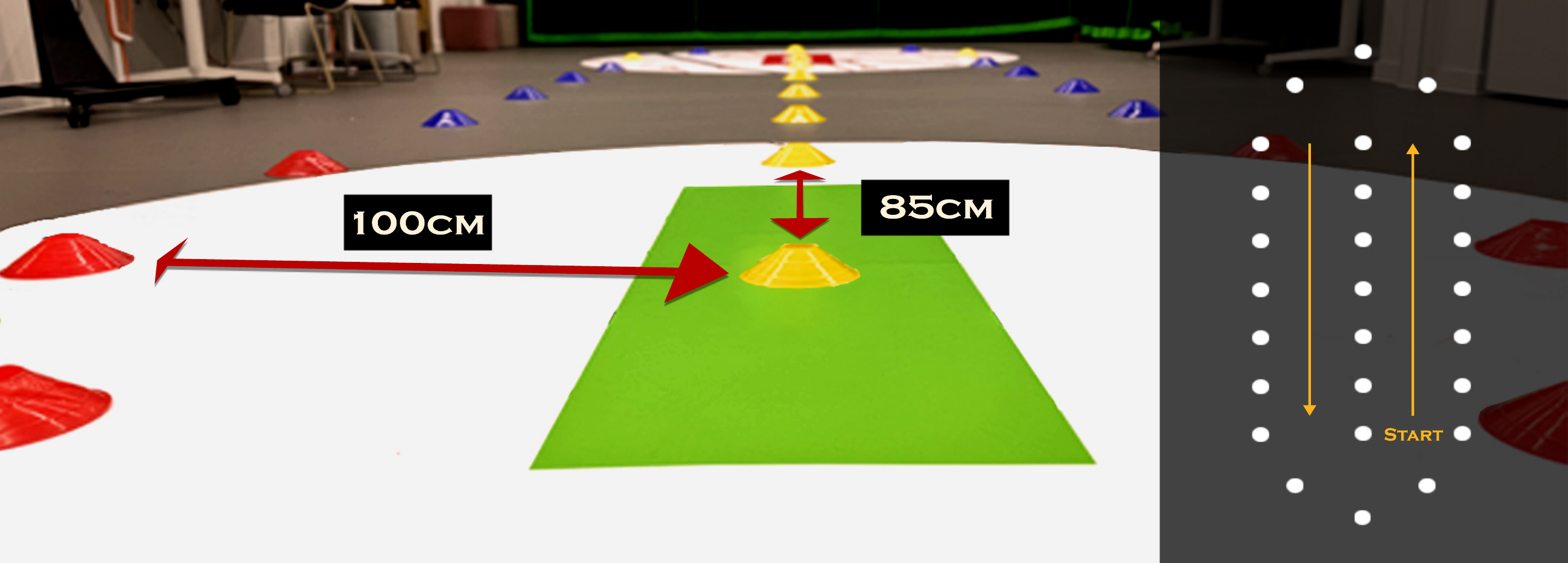}
    
  \caption{Participants walking path during the experiment. Similar to prior work \cite{10.1145/2628363.2628382}, we placed cones to guide the path that participants were required to traverse. Each cone was spaced approximately 85 cm apart, and the path was about one metre wide, as illustrated using the red arrows in the left image. Participants completed tasks in two conditions: encumbered by carrying bags and unencumbered. They were instructed to walk all the way to the end of the path, return via the designated return route, and continue walking additional laps until all required selections for the presented condition were completed.}
  \Description{}
  \label{fig:path}
\end{figure*}

We conducted the experiment in a quiet, spacious laboratory on the university campus, measuring approximately 10 m $\times$ 5 m (see Figure \ref{fig:path}). We aimed to explore the impact of encumbrance on target selection using gaze alone, or supported by touch, on handheld mobile devices and compare that to the touch input as a baseline. Participants completed all selection tasks while a) encumbered by carrying bags in both hands or b) not encumbered. Following practices from prior work \cite{10.1145/3544548.3580871}, they were instructed to walk as they would naturally without pausing. 
The experimenter observed participants during walking to identify instances of slowing or stopping; however, such instances occurred only a few times, and participants were instructed to maintain a consistent pace. 
The Experiment received ethical approval from our institution [Approval number: 300240212]. We compensated participants with a £10 e-shop voucher.

\subsection{Study Design} \label{sec:design}
This within-subjects experiment had two main independent variables:

\begin{itemize}
    \item \textbf{Input Modality} 
    \begin{itemize}
        \item \textit{Gaze}.
        \item Gaze with visual feedback (\textit{\gazeF}).
        \item \textit{GazeTouch}: gaze is used for pointing and touch for selecting.
        \item GazeTouch with visual feedback (\textit{\gazetouchF}).
        \item Touch one-handed (\textit{\touchOne}).
        \item Touch two-handed (\textit{\touchTwo}).  
    \end{itemize}
    
For all Gaze and GazeTouch conditions, participants held the device in their preferred hand. They used their thumb to touch for GazeTouch. Gaze and GazeTouch levels with visual feedback (\gazeF and \gazetouchF) provided participants with visual feedback described in Section \ref{sec:visualF} (See Figure \ref{fig:screenshot}). 
For the touch input, we used two common hand postures \cite{10.1145/2628363.2628382}: 1) when using the preferred hand to hold the phone and interact using the thumb (\touchOne), and 2) when the participant held the phone in the non-dominant hand while using the index finger of the other hand to interact (\touchTwo).
    \item \textbf{Encumbrance Level} 
    \begin{itemize}
    \item Walking encumbered.
    \item Walking unencumbered

    \end{itemize}
    Following practices from prior work, the \textit{walking encumbered} required participants to carry two typical supermarket carrier bags, with one bag in each hand. Each bag was filled with three 500 ml water bottles, resulting in a total mass of approximately 1.5 kg per bag. The chosen weight, which was motivated by prior work~\cite{10.1145/2556288.2557312}, was designed to replicate the effects of holding a realistic object while minimising fatigue and strain on participants. The walking unencumbered required participants to walk and interact with the mobile device without carrying bags in their hands.
\end{itemize}

\subsection{Participants}
We recruited $24$ participants for the experiment ($7$ male, $16$ female, $1$ prefer self-describe; ages 19-49, $M=26.42$, $SD=6.5$). While two participants wore glasses during the experiment and one wore contact lenses, 10 participants reported wearing glasses daily, while another two reported wearing contact lenses. Out of the $24$ participants, 14 reported vision-related conditions: three had astigmatism, four were farsighted, and four were nearsighted. One participant reported both nearsightedness and astigmatism, while another reported all three conditions: nearsightedness, farsightedness, and astigmatism. One participant reported one eye being nearsighted and the other farsighted. Out of the $24$ participants, $19$ indicated that their right hand is preferred when holding their mobile devices. While $21$ selected their right hand as their dominant hand, two selected their left hand, and one reported no clear dominance. Participants generally reported that they \say{sometimes} to \say{often} use their phone while walking ($Mdn= 3.5$, $Q1=3$,$Q3=4$), reflecting a relatively common behaviour among them (0= never; 5= always). Participants reported limited to moderate experience ($Mdn=2$, $Q1=1$, $Q3=3$) with eye tracking (0= no experience; 5= very experienced).

\subsection{Apparatus}
In this study, we used an iPhone 16 Plus with a front-facing camera with a 12MP sensor and an f/1.9 aperture. Touch input was enabled using the built-in tap gesture using the touchscreen of Apple devices, while gaze input was enabled using the Eyedid tracking library \cite{seeso}. Eyedid is a software-based eye-tracking system that provides gaze estimates as x and y coordinates, utilising the RGB images collected from the phone's front-facing camera at 30 frames per second. 

\begin{figure}[!t]
  \centering
  \includegraphics[width=.8\linewidth]{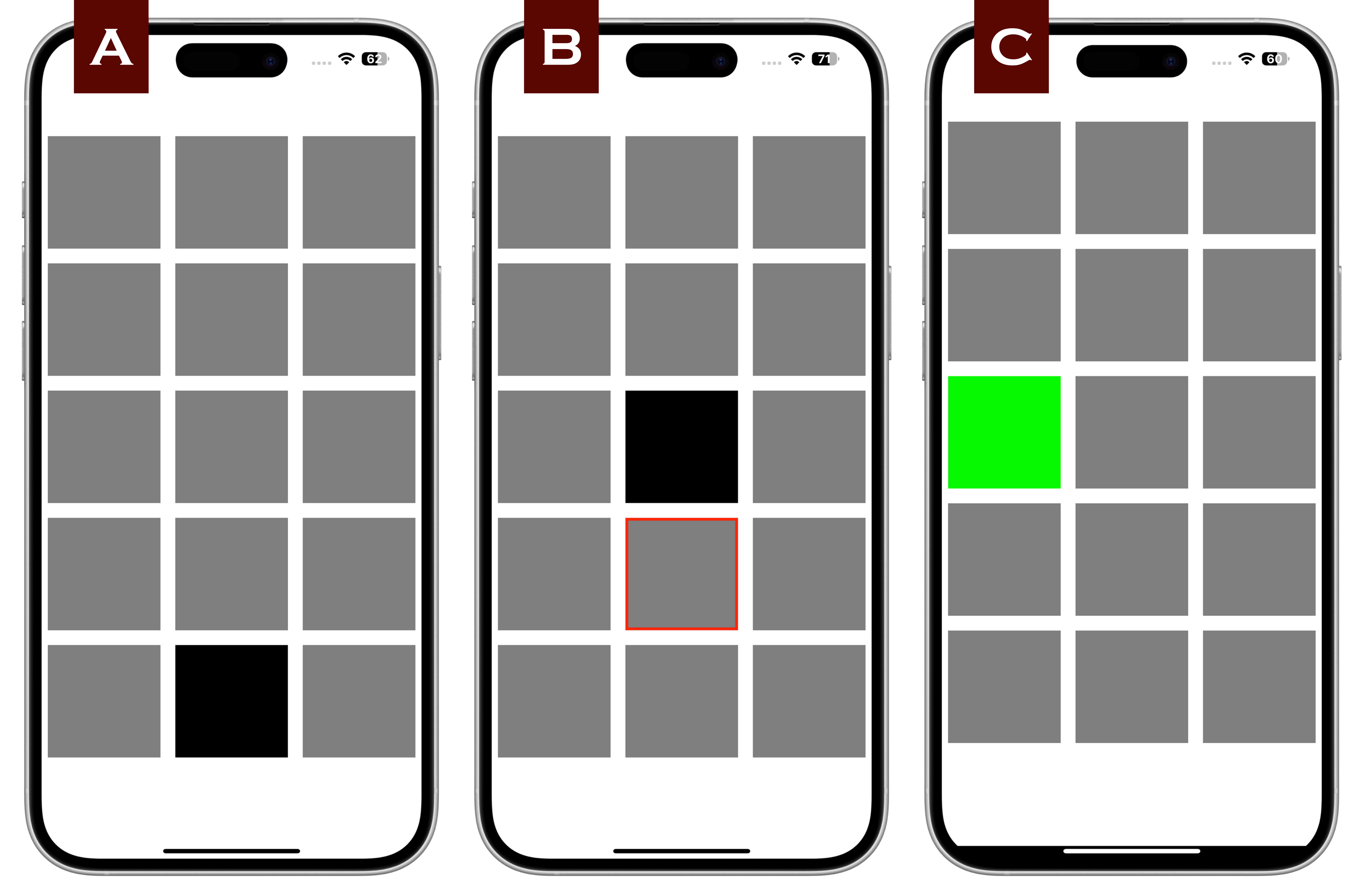}
    
  \caption{In each encumbrance level, participants had to complete nine selection tasks per input modality. In particular, they had to select the three targets at the top, three in the middle, and three in the bottom region. The target to be selected is highlighted in black \textbf{(A)}. Visual feedback using red strokes surrounding the target the participants are looking at was provided in the Gaze (feedback) and GazeTouch (feedback) conditions \textbf{(B)}. Once the correct target is selected, it changes its colour to green \textbf{(C)}.}
  \Description{}
  \label{fig:screenshot}
\end{figure}

Following the recommendation from prior work on target sizes for dwell interfaces on handheld mobile devices \cite{10.1145/3706598.3713092}, each target was sized at 4\textdegree, computed at a distance of 30 cm between the participant and the device. The distance used to determine the computed on-screen target size is inspired by prior work on common distances based on posture, where researchers suggested that users hold their phone at a range of between 30 and 40 cm when walking, sitting, and standing \cite{10.1145/3025453.3025794}. Such a design allowed us to fit three targets in each row and five targets across the screen vertically. We maintained the same spacing between targets vertically and horizontally (see Figure \ref{fig:screenshot}).

\subsection{Procedure}
After an initial briefing, signing the consent form, and completing the demographic questionnaire, we calibrated the eye tracker using the front-facing camera for each participant before the experiment began. The calibration procedure was provided by EyeDid \cite{seeso}, the eye-tracking library used in the experiment. Participants completed the experiment in two blocks, with each block corresponding to a different level of encumbrance. Half of the participants started with walking unencumbered, followed by walking encumbered, while the other half completed the blocks in the reverse order.
\textbf{In each block,} the task was to select targets from a three-by-five grid as accurately and quickly as possible using six input modalities (see Section \ref{sec:design}). While the prototype included 15 targets, time constraints to account for participant fatigue required us to limit the number of selectable targets per input modality to nine: specifically, the three targets at the top, three in the middle, and three at the bottom (see Figure \ref{fig:screenshot}).
Participants performed nine selections using one input modality at a time, resulting in 54 selections per block.
The order of input modalities was counterbalanced across participants using a Latin square design. Before completing the tasks in each block, participants underwent a training phase to familiarise themselves with the input modalities, during which they selected a target once with each modality.
\textbf{For each input modality,} we presented the nine targets to be selected, one at a time, in sequence (serially), rather than as discrete trials to capture the essence of a real user selecting a real object based on their interests \cite{10.1145/332040.332445}. There was a random interval ranging from 0.5 to 1.5 seconds between a selection and the next target being highlighted for selection to negate any rhythm created between the participant’s walking and onscreen targets \cite{10.1145/2556288.2557312}.
For each selection, we displayed all targets in a grey background colour and randomly highlighted one for selection in a black background. No Targets repeated twice in a row for selection. Once the correct target is selected, its colour changes to green and the next target is shown. All targets were selectable, and all wrong selections were logged.
Participants were instructed to complete the selection as soon as the highlighted target was visible. If participants failed to select the correct target within 5 seconds, it would result in a timeout \cite{10.1145/2168556.2168601, Fernandes2025}, ending the trial. 
\textbf{After completing the selection for each input modality within a block,} participants filled out the NASA-TLX questionnaire to assess their perceived workload. Additionally, they responded to Likert scale questions regarding usability aspects.
\textbf{After each block,} participants ranked their preference for input modalities based on their experience in each encumbrance level. Participants completed a total of 108 selections (2 encumbrance levels x 6 input modalities x 9 regions).
\textbf{After completing the two blocks}, we conducted a semi-structured interview to reflect on the participants' ranking decision of input modalities based on encumbrance level. The experiment lasted for approximately an hour.

\subsection{Measures} \label{sec:measures}
We measured selection time, completion rate, and error counts.
We defined selection time as the time from the moment the target is displayed until the correct target is selected.
The completion rate was calculated as the percentage of successful target selections within the given time, as we set a timeout of 5 seconds to end a trial if the highlighted target was not selected successfully \cite{10.1145/2168556.2168601, Fernandes2025}. 
We counted the number of times participants selected incorrect targets before selecting the correct one as the error count. The error counts also include instances when participants' tapping does not result in any selection, particularly when touch is involved. We averaged all measures per input modality per encumberance level. 
Besides the quantitative measures, we measured perceived usability and cognitive workload using Likert scale questions and NASA-TLX \cite{doi:10.1177/154193120605000909}, respectively. We also asked participants to rank their preferences for input modalities and collected qualitative feedback through semi-structured interviews to gain insight into their rankings and preferences.

\subsection{Limitations}
Our findings are based on designing targets for dwell selection using the recommended size of 4° from prior work \cite{10.1145/3706598.3713092}. The targets were measured at a viewing distance of 30 cm between the participant and the device. Previous work has reported that users typically hold their phones at distances of 30-40 cm while walking \cite{10.1145/3025453.3025794}. To represent the closest and most conservative interaction distance within this range, we adopted 30 cm as the reference distance for our design. This distance also allowed us to fit three targets comfortably in a single row on the screen. However, the mobile context may result in the distance to vary as prior work showed \cite{10.1145/3025453.3025794, 10.1145/3706598.3713092}. 
We expect that such variation may impact the gaze-based input modalities.
In addition, the chosen target sizes were used across all conditions for consistency. While such a decision ensured fair comparison, we expect some positive/negative impact on the performance of touch and gaze inputs. 
Future work can further optimise the gaze interface beyond just changing the target shape, for example, by adapting the target size to maximise accuracy at a given distance. Recent work, Mind the Gaze \cite{}, showed that dwell interfaces could benefit from adjusting or adapting target sizes to account for different face-to-screen distances when interacting with handheld mobile devices. 
In our experiment, we attempted to quantify the impact of walking while being encumbered on gaze input and compare that with touch input as a baseline. Thus, we intentionally chose an indoor environment with minimal distractions and controlled lighting conditions to investigate the effect of encumbrance in isolation. While the real world is more complex, the participants had no obstacles to avoid during the interaction, which might have affected the results if they had been present.

\begin{figure*}[!t]
  \centering
  \includegraphics[width=.9\linewidth]{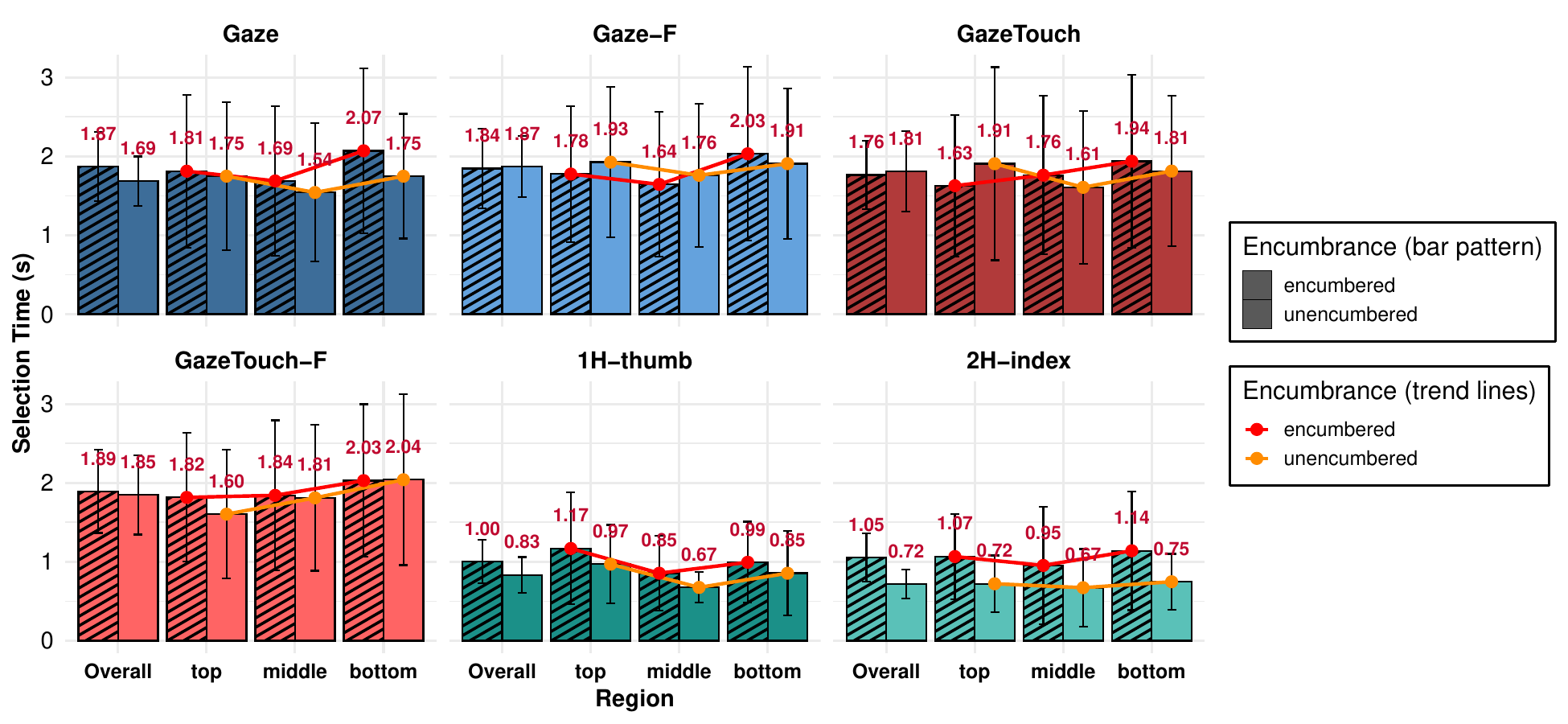}
    
  \caption{Encumbrance significantly impacted the overall selection time regardless of the region for \touchOne and \touchTwo, where the selection time was significantly slower when the participants were encumbered compared to when they were unencumbered. Selecting targets from the middle regions was significantly faster than the top regions when using \touchOne while encumbered. When unencumbered, selection time differed significantly across regions for \touchOne. The red and orange lines show the trend in encumbrance across regions. Error bars represent standard deviations. }
  \Description{}
  \label{fig:selection_time}
\end{figure*} 

\section{Results}
 We conducted statistical tests using a two-way repeated-measures ANOVA, unless otherwise noted. When there is an interaction between the independent variables, we run a separate one-way ANOVA to examine the impact of encumbrance on each input modality. We employed the Aligned Ranked Transform (ART) to transform data that was not normally distributed \cite{10.1145/1978942.1978963}, as determined by the Shapiro-Wilk test. P-values were corrected using the Bonferroni correction method to account for multiple comparisons for post-hoc analysis. 
 Because ANOVA tests that show no significant effects cannot be used to determine lack of effects,  for selection time, completion rate, and error counts, we used Bayesian methods \cite{10.1145/3613904.3642712, 10.1145/3706598.3713492} to quantify uncertainty and facilitate future work. 
 We tested our hypotheses regarding the presence of the encumbrance effect across input modalities, such that encumbrance increases selection time, reduces completion rate, and increases error counts. For each input modality, we modelled Participant ID as a random effect with a fixed effect of encumbrance.
 Following a similar approach to prior work \cite{10.1145/3706598.3713492}, we fit Models using the brms package in R (Stan) with regularising priors, assessed the convergence with Rhat ($<1.01$) and Effective Sampling Size ($>1000$) \cite{JSSv080i01}, and we report posterior means with 95\% credible intervals. 
 Instead of p-values, we interpret Bayes Factors (BF) as measures of evidence, which compares the liklihood the observed data under the proposed model over the null condition \cite{wagenmakers2011why, russo2003statistics, 10.1145/3706598.3713492}: $BF<3$ indicates anecdotal evidence, $3 - 10$ substantial, $10 - 30$ strong, $30 - 100$ very strong, and $>100$ extreme evidence in favor of the hypothesis, whereas $BF < 1$ indicate evidence supporting the null hypothesis (i.e., no difference between conditions), with smaller values representing stronger evidence. We used Bayesian statistics solely to quantify uncertainty; therefore, no statistical significance should be inferred from the results. 
 We also report posterior probabilities, interpreted using approaches from prior work \cite{10.1145/3706598.3713492, 10.1145/3613904.3642919}, where values > 90\% support the hypothesis (positive effect of encumbrance), 60–90\% suggest moderate evidence, 40–60\% indicate uncertainty, 10–40\% suggest weak evidence against, and < 10\% provide strong evidence for a negative effect. We report the posterior mean (PM), the estimated error (SE), and the lower and upper bounds of the 95\% credible interval (CI=[lower, upper]). 
 Out of the $2592$ data points collected from $24$ participants, data from 3 trials were lost due to hardware error, resulting in a total of $2589$ data points used for analysis ($24$ participants $\times$ $6$ input modalities $\times$ $2$ encumbrance levels $\times$ $9$ selections, 3 per region - $3$).

\subsection{Selection Time}
There was a significant main effect of \textit{Encumbrance Level} on selection time (F\textsubscript{1,253} = 5.38, $p<.05$). Post hoc analysis revealed that participants completed the selections significantly faster ($p<.05$) when unencumbered ($M=1.41$, $SD=.94$) than when encumbered ($M=1.52$, $SD=.94$).
There was a significant main effect of \textit{Input Modality} on selection time (F\textsubscript{5,253} = 99.78, $p<.0001$). \touchOne ($M=.92s$, $SD=.53s$) and \touchTwo ($M=.88s$, $SD=.59s$) were significantly faster ($<.0001$) than Gaze ($M=1.77s$, $SD=.94s$), \gazeF ($M=1.84s$, $SD=.95s$), GazeTouch ($M=1.77s$, $SD=1.03s$), and \gazetouchF ($M=1.85s$, $SD=.93s$).  There was no interaction effect between \textit{Input Modality} and \textit{Encumbrance Level} on selection time, F\textsubscript{5,253} = 2.09, $p=.07$.

\subsubsection{Impact of Encumbrance on Input Modalities}
To determine how selection time under each modality is impacted by encumbrance, we examined the effect of \textit{Encumbrance Level} on selection time across input modalities. Figure \ref{fig:selection_time} shows the mean selection time for each input modality in the two encumbrance levels. We employed Bayesian regression to model the effect of encumbrance on selection time across input modalities in cases where the ANOVA revealed no significant differences. The selection time follows a shifted Log-normal distribution. 

We found no evidence that the mean selection time differed significantly across encumbrance levels for Gaze (F\textsubscript{1,23} = 4.06, $p>.05$), \gazeF (F\textsubscript{1,23} = 0.82, $p>.05$), GazeTouch (F\textsubscript{1,23} = 0.11, $p>.05$) and \gazetouchF (F\textsubscript{1,23} = .11, $p>.05$).
The Bayesian models suggested that walking encumbered has a $96.11\%$ probability of leading to longer selection time ($PM=.16, SE=0.10$, $CI=[-.02, 0.37]$) for Gaze, and a $27.95\%$ probability for \gazeF ($PM=-.06$, $SE=.10$, $CI=[-.28,.14]$). 
The Bayes factor of $1.13$ for Gaze indicated weak support for encumbrance leading to longer selection time. The Bayes factor of $0.25$ for \gazeF supports the null hypothesis of no difference in selection time. 
The probability that walking encumbered increased selection time was $44.38\%$, with a coefficient ($PM=-.01$, $SE=.10$, $CI=[-.20,.18]$) for GazeTouch, and $59.75\%$ ($PM=.03$, $SE=.13$, $CI=[-.22,.29]$) for \gazetouchF.
The Bayes Factors of $0.22$ for GazeTouch, and $0.27$ for \gazetouchF provided evidence in favour of the null model; i.e., no support for encumbrance leading to longer selection time.
For \touchOne, the mean selection time differed statistically significantly between encumbrance levels (F\textsubscript{1,23} = 7.85, $p<.05$). Participants performed selection significantly faster ($p<.05$) when unencumbered ($M=.83s$, $SD=.23s$) compared to when encumbered ($M=1s$, $SD=.28s$). 
Similarly, the mean selection time differed statistically significantly between encumbrance levels for \touchTwo (F\textsubscript{1,23} = 31.68, $p<.0001$). Encumbrance caused selection time ($M=1.05s$, $SD=.30s$) to be significantly longer ($<.0001$) compared to when unencumbered ($M=.72s$, $SD=.18s$).

\observation{Participants selected targets significantly slower with \touchOne and \touchTwo while encumbered than while unencumbered. Using Gaze when encumbered showed weak evidence of an increase in selection time, while \gazeF, GazeTouch, and \gazetouchF showed comparable selection times when performing selections while encumbered and unencumbered.}

\subsubsection{Impact of Target Region on Input Modalities}
We explored the effect of the target region (top, middle, and bottom) on selection time across input modalities at each encumbrance level. For each modality, a repeated-measures ANOVA was conducted with target region as the within-subjects factor. See Figure \ref{fig:selection_time} for the descriptive statistics. 

The statistical tests did not reveal significant differences in mean selection time ($p>.05$) between the three regions across all gaze-based inputs while encumbered (Gaze: F\textsubscript{2,44.93} = 2.27; \gazeF: F\textsubscript{2,45.39} = 2.55; GazeTouch: F\textsubscript{2,45.50} = .39; \gazetouchF: F\textsubscript{2,43.76} = .57) and also while unencumbered (Gaze: F\textsubscript{2,46} = 1.79; \gazeF: F\textsubscript{2,45.07} = .38; GazeTouch: F\textsubscript{2,44.88} = .90; \gazetouchF: F\textsubscript{2,45.26} = 2.57). 
On the other hand, for \touchOne, when encumbered, the mean selection time differed significantly between regions (F\textsubscript{2,46} = 8.93, $p<.001$). Selecting targets from middle region ($M=.86s$, $SD=.28s$) was significantly faster ($p<.0005$) than selecting targets from top region ($M=1.16s$, $SD=.44s$). We found no significant differences ($p>.05$) between other pairs. 
Similarly, when unencumbered, the mean selection time differed significantly between regions (F\textsubscript{2,46} = 19.26, $p<.0001$). Post hoc analysis revealed significant differences between all regions. 
Selecting a target from the middle region ($M=.68s$, $SD=.15s$, $p<.0001$) was significantly faster compared to the top ($M=.97s$, $SD=.33s$, $p<.0001$) and bottom ($M=.86s$, $SD=.36s$, $p<.005$) regions. Selection from the bottom regions was significantly faster ($p<.05$) than selection from the top region.
When using \touchTwo while encumbered and unencumbered, the tests revealed significant differences in selection time ($p<.05$) across regions (encumbered: F\textsubscript{2,46} = 3.44, $p<.05$; unencumbered: F\textsubscript{2,46} = 3.63, $p<.05$). However, post hoc analysis did not reveal significant differences ($p>.05$).

\observation{For \touchOne, participants selected targets in the middle region significantly faster than in the top region while encumbered, and significantly faster than in the top and bottom regions while unencumbered. Selection from the bottom region was also significantly faster than from the top region when unencumbered.}

\begin{figure}[!t]
  \centering
  \includegraphics[width=\linewidth]{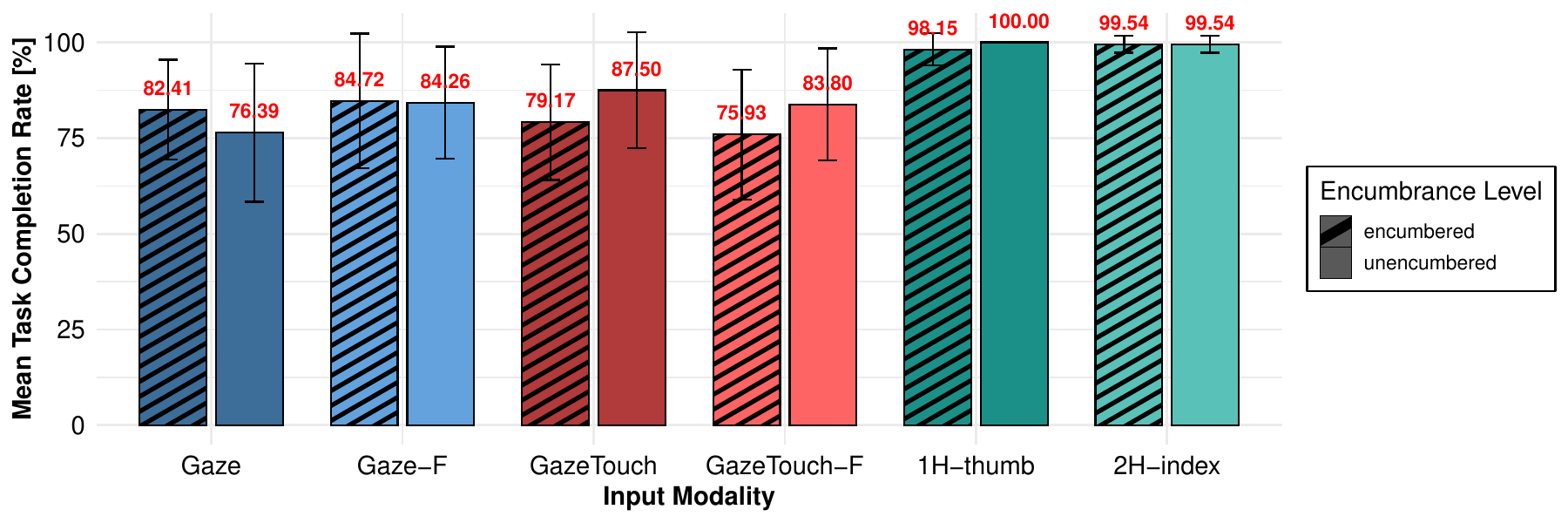}
    
  \caption{We calculated the percentage for successful selection of targets for each input modality while walking unencumbered and encumbered, as the completion rate. We found that encumbrance significantly reduced completion rates for GazeTouch, \gazetouchF, and \touchOne input modalities. The error bars represent the standard deviations.}
  \Description{}
  \label{fig:selection_Accuracy}
\end{figure} 

\subsection{Task Completion Rate}
The statistical test revealed a significant interaction effect between \textit{Input Modality} and \textit{Encumbrance Level} on completion rate (F\textsubscript{5,253} = 3.71, $p<.005$). Due to interaction, we explored the effect of \textit{Encumbrance Level} on completion rate across input modalities. We used Bayesian regression to model the effect of encumbrance on selection time across input modalities when the ANOVA test did not reveal significant differences. The task completion rate was modelled using a binomial likelihood, reflecting the number of completed selections out of nine trials per input modality for each encumbrance level. See Figure \ref{fig:selection_Accuracy} for descriptive statistics. 

The statistical tests did not reveal that the task completion rate differed significantly across encumbrance levels for Gaze (F\textsubscript{1,23} = 1.55, $p>.05$), or \gazeF (F\textsubscript{1,23} = .50, $p>.05$).
The Bayesian models estimated that walking encumbered has a $5.54\%$ probability of lowering the task completion rate ($PM=.39$, $SE=.24$, $CI=[-.10,.87]$) for Gaze, and a $44.53\%$ probability for \gazeF ($PM=.04$, $SE=.28$, $CI=[-.51,.59]$). The Bayes Factor of $2.09$ for Gaze indicated anecdotal evidence, while the Bayes Factor of $.72$ for \gazeF support the null hypothesis of no difference in task completion rate.

The statistical tests showed that the completion rate differed statistically significantly between encumbrance levels when using GazeTouch (F\textsubscript{1,23} = 7.70, $p<.05$) and \gazetouchF (F\textsubscript{1,23} = 6.24, $p<.05$). 
When using GazeTouch, participants completed significantly ($p<.05$) less tasks when encumbered ($M=79.2\%$, $SD=15.1\%$) than when unencumbered ($M=87.5\%$, $SD=15.1\%$). Similarly, when using \gazetouchF, they completed significantly ($p<.01$) less tasks when encumbered ($M=75.9\%$, $SD=16.9\%$) compared to when unencumbered ($M=83.8\%$, $SD=14.6\%$).

The statistical test determined that the completion rate differed significantly between encumbrance levels when using \touchOne (F\textsubscript{1,23} = 4.6, $p<.05$). Completion rate was significantly reduced ($p<.05$) with encumbrance ($M=98.1\%$, $SD=4.23\%$) compared to when participants were unencumbered ($M=100\%$, $SD=0\%$). 
While we found no evidence that the completion rate differed significantly across encumbrance Levels when using \touchTwo (F\textsubscript{1,23} = 0, $p>.05$), the probability that walking encumbered lowered task completion rate was $49.9\%$, indicating uncertainty ($PM=-.004$, $SE=1.61$, $CI=[-3.19,3.24$]). The Bayes Factor of $3.60$ suggested that encumbrance has a substantial impact on task completion rate.

\observation{Encumbrance significantly reduced task completion rates for GazeTouch, \gazetouchF, and \touchOne, where participants completed significantly less tasks when encumbered. For Gaze and \gazeF, we found anecdotal to no evidence for the effect of encumbrance on task completion rate; for \touchTwo, the results suggest that encumbrance has a substantial effect on task completion rate.}

\subsubsection{Impact of Target Region on Input Modalities}
We explored the impact of the target region (top, middle, or bottom) on the completion rate for each input modality across encumbrance levels. For each modality, a repeated-measures ANOVA was conducted with target region as the within-subjects factor.

The statistical tests did not reveal significant differences in completion rate ($p>.05$) across regions when using Gaze (F\textsubscript{2,46} = .33), \gazeF (F\textsubscript{2,46} = 2.04), \touchOne (F\textsubscript{2,46} = 1.87), and \touchTwo (F\textsubscript{2,46} = 1) while encumbered. 
When unencumbered, we found no evidence of significant differences in completion rate ($p>.05$) across regions, when using Gaze (F\textsubscript{2,46} = .30), \gazeF (F\textsubscript{2,46} = 1.33), GazeTouch (F\textsubscript{2,46} = 1.86), \gazetouchF (F\textsubscript{2,46} = .76), \touchTwo (F\textsubscript{2,46} = 1). For \touchOne when used unencumbered, we could not perform an ANOVA to compare regions because there was no variability in completion rates.
On the other hand, when using GazeTouch and \gazetouchF when encumbered, the statistical test revealed that the task completion rate differed significantly between regions when using both (GazeTouch: F\textsubscript{2,46} = 3.62, $p<.05$; \gazetouchF: F\textsubscript{2,46} = 5.27, $p<.01$). Participants completed significantly ($p<.05$) more selections in the top regions (GazeTouch: $M=87.5\%$, $SD=16.5\%$; \gazetouchF: $M=84.7\%$, $SD=26.0\%$) compared to the bottom region (GazeTouch: $M=69.4\%$, $SD=27.7\%$; \gazetouchF: $M=61.1\%$, $SD=33.6\%$). 

\observation{Participants completed significantly more selection tasks in the top region than the bottom region when using GazeTouch and \gazetouchF while encumbered. For Gaze, \gazeF, \touchOne and \touchTwo, task we found no evidence for significant differences in task completion rate across the screen region.}

\begin{figure}[!t]
  \centering
  \includegraphics[width=\linewidth]{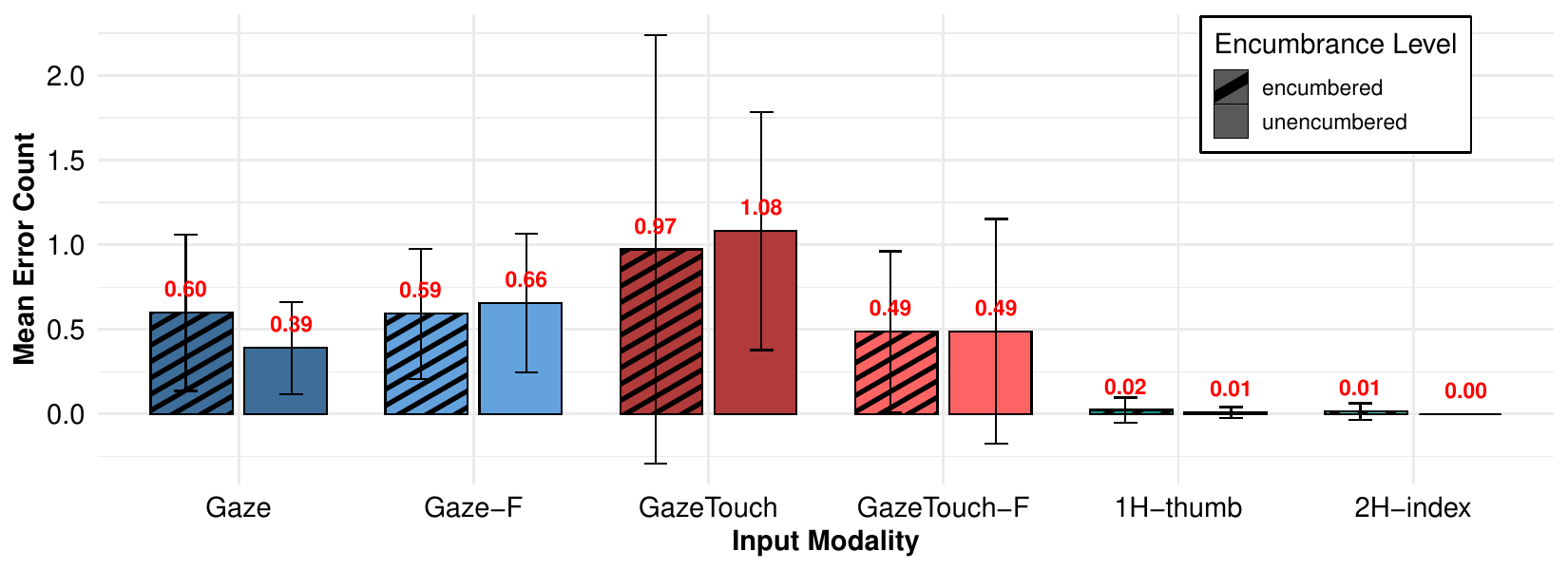}
    
  \caption{We counted the number of times participants selected incorrect targets before selecting the correct one. While error counts varied across modalities due to encumbrance, we found no statistically significant differences. The error bars represent the standard deviation.}
  \Description{}
  \label{fig:error_count}
\end{figure} 

\subsection{Error Counts}
The statistical tests revealed a significant interaction between \textit{Input Modality} and \textit{Encumbrance Level} on error counts (F\textsubscript{5,253} = 2.79, $p=.05$). Due to the interaction effect, we explored the impact of encumbrance on error counts across input modalities. For each modality, we also modelled the error counts out of the total number of trials using a Poisson to estimate the effect of encumbrance. See Figure \ref{fig:error_count} for descriptive statistics. 

The statistical tests did not reveal that the error counts differed significantly ($p>.05$) across encumbrance levels for Gaze (F\textsubscript{1,23} = 4.19), \gazeF (F\textsubscript{1,23} = .39), GazeTouch (F\textsubscript{1,23} = 3.11), and for \gazetouchF (F\textsubscript{1,23} = 1.51).
The Bayesian models estimated that walking encumbered has a $99.56\%$ probability of leading to more errors ($PM=.41$, $SE=.16$, $CI=[.10,.72]$) when using Gaze, and a probability of $22.78\%$ when using \gazeF ($PM=-.10$, $SE=.13$, $CI=[-.36,.16]$) 
The Bayes Factor of $11.88$ indicated strong evidence for differences when using Gaze, and the Bayes Factor of $.42$ supports the null hypothesis of no difference when using \gazeF.
The models estimated that walking encumbered has a $34.36\%$ probability of leading to more errors when using GazeTouch, with a coefficient ($PM=-.04$, $SE=.11\%$, $CI=[-.25, .16]$), and a probability of $34.04\%$ when using \gazetouchF ($PM=-.06$, $SE=.16$, $CI=[-.37,.24]$). 
The Bayes Factor of $.28$ when using GazeTouch, and $.41$ when using \gazetouchF supported the null hypothesis of no differences in errors due to encumbrance.
The statistical tests showed no evidence that the error counts differed significantly across encumbrance levels for \touchOne (F\textsubscript{1,23} = .37, $p>.05$), and for \touchTwo (F\textsubscript{1,23} = 2.09, $p>.05$).
The Bayesian models estimated that waking encumbered has $88.99\%$ probability of leading to more errors when using \touchOne ($PM=1.06$, $SE=.91$, $CI=[-.55,3.05]$), and a probability of $98.6\%$ when using \touchTwo ($PM=6.89$, $SE=7.13$, $CI=[.37,24.25]$)
The Bayes Factor of $4.43$ indicated substantial evidence for differences when using \touchOne, and the Bayes Factor of $40.63$ when using \touchTwo indicated very strong evidence for differences in error counts due to encumbrance.

\observation{While we found no evidence for significant differences in error counts due to encumbrance across the input modalities, Bayesian analysis suggested that walking encumbered increased error counts for Gaze with strong evidence, \touchOne with substantial evidence, and  \touchTwo with very strong evidence.}

\subsubsection{Impact of Target Region on Input Modalities}
We explored the impact of the target region (top, middle, or bottom) on error counts for each input modality across encumbrance levels. For each modality, we conducted repeated-measure ANOVA with the target region as the within-subject factor.

When walking encumbered, the statistical tests did not show that the error counts differed significantly ($p>.05$) between regions across all input modalities (Gaze: F\textsubscript{2,44.88} = .71; \gazeF: F\textsubscript{2,45.51} = .70; GazeTouch: F\textsubscript{2,45.52} = .20; \gazetouchF: F\textsubscript{2,44.45} = .49; \touchOne: F\textsubscript{2,46} = 1.04; \touchTwo: F\textsubscript{2,46} = 1.53).
Similarly, when walking unencumbered, there was no evidence for statistically significant differences ($p>.05$) in error counts between regions across all input modalities (Gaze: F\textsubscript{2,46} = 1.24; \gazeF: F\textsubscript{2,45.03} = .54; GazeTouch: F\textsubscript{2,44.85} = 2.72; \gazetouchF: F\textsubscript{2,45.25} = 1.14; \touchOne: F\textsubscript{2,46} = .5). 
For \touchTwo, we could not perform ANOVA because the data showed no variability in error counts when walking unencumbered, with zero errors across all regions.

\begin{table*}[t!]
\centering
\caption{The table highlights significant differences in perceived usability, showing for each aspect which input modality was significantly impacted by encumbrance. We report the median ratings, both when encumbered and unencumbered, along with the Friedman test results. The interquartile range (IQR) values are given in brackets. Non-significant results are included in Table \ref{tab_likert_significance_all}, in the appendices for clarity. The reported rating shows that touch-involved inputs are mostly impacted by encumbrance.}
\label{tab_likert_significance}
\resizebox{.6\textwidth}{!}{%
\begin{tabular}{llllll}
                                                                                                                &                                              & \multicolumn{2}{c}{\textbf{Encumbrance Level}}                                        & \multicolumn{2}{c}{\textbf{Friedman}}                                            \\ 
\multicolumn{1}{l}{\textbf{Usability Aspect}}                                                                 & \multicolumn{1}{l}{\textbf{Input Modality}} & \multicolumn{1}{l}{\textbf{Encumbered}} & \multicolumn{1}{l}{\textbf{Unencumbered}} & \multicolumn{1}{l}{\textbf{$\chi\textsuperscript{2}(1)$}} & \multicolumn{1}{l}{\textbf{p\textless{}}} \\ \hline \hline
\multicolumn{1}{l}{\multirow{2}{*}{Natural}}                                                                  & \multicolumn{1}{l}{\gazetouchF}    & \multicolumn{1}{l}{3 (1)}       & \multicolumn{1}{l}{3 (1)}         & \multicolumn{1}{l}{4.76}           & \multicolumn{1}{l}{.05}                   \\ 
\multicolumn{1}{l}{}                                                                                          & \multicolumn{1}{l}{\touchTwo}        & \multicolumn{1}{l}{4 (2)}       & \multicolumn{1}{l}{4 (1)}         & \multicolumn{1}{l}{15}             & \multicolumn{1}{l}{.0005}                 \\ \hline
\multicolumn{1}{l}{\multirow{2}{*}{Fast}}                                                                     & \multicolumn{1}{l}{\touchOne}        & \multicolumn{1}{l}{4 (1.25)}    & \multicolumn{1}{l}{5 (1)}         & \multicolumn{1}{l}{5.4}            & \multicolumn{1}{l}{.05}                   \\ 
\multicolumn{1}{l}{}                                                                                          & \multicolumn{1}{l}{\touchTwo}        & \multicolumn{1}{l}{4 (1)}       & \multicolumn{1}{l}{5 (1)}         & \multicolumn{1}{l}{13.24}          & \multicolumn{1}{l}{.0005}                 \\ \hline
\multicolumn{1}{l}{\multirow{2}{*}{Enjoyable}}                                                                & \multicolumn{1}{l}{\touchOne}        & \multicolumn{1}{l}{3 (1.25)}    & \multicolumn{1}{l}{4 (2)}         & \multicolumn{1}{l}{8.05}           & \multicolumn{1}{l}{.005}                  \\ 
\multicolumn{1}{l}{}                                                                                          & \multicolumn{1}{l}{\touchTwo}        & \multicolumn{1}{l}{3 (1)}       & \multicolumn{1}{l}{4 (1)}         & \multicolumn{1}{l}{9.94}           & \multicolumn{1}{l}{.005}                  \\ \hline
\multicolumn{1}{l}{\multirow{3}{*}{Easy to use}}                                                              & \multicolumn{1}{l}{\gazetouchF}    & \multicolumn{1}{l}{4 (2)}       & \multicolumn{1}{l}{4 (2)}         & \multicolumn{1}{l}{4}              & \multicolumn{1}{l}{.05}                   \\  
\multicolumn{1}{l}{}                                                                                          & \multicolumn{1}{l}{\touchOne}        & \multicolumn{1}{l}{4 (2)}       & \multicolumn{1}{l}{5 (1)}         & \multicolumn{1}{l}{9}              & \multicolumn{1}{l}{.005}                  \\ 
\multicolumn{1}{l}{}                                                                                          & \multicolumn{1}{l}{\touchTwo}        & \multicolumn{1}{l}{3 (2)}       & \multicolumn{1}{l}{5 (1)}         & \multicolumn{1}{l}{14.22}          & \multicolumn{1}{l}{.0005}                 \\ \hline
\multicolumn{1}{l}{\multirow{3}{*}{Accurate}}                                                                 & \multicolumn{1}{l}{GazeTouch}               & \multicolumn{1}{l}{3 (1.25)}    & \multicolumn{1}{l}{4 (1)}         & \multicolumn{1}{l}{5.44}           & \multicolumn{1}{l}{.05}                   \\ 
\multicolumn{1}{l}{}                                                                                          & \multicolumn{1}{l}{\gazetouchF}    & \multicolumn{1}{l}{3 (2)}       & \multicolumn{1}{l}{4 (1)}         & \multicolumn{1}{l}{5.4}            & \multicolumn{1}{l}{.05}                   \\ 
\multicolumn{1}{l}{}                                                                                          & \multicolumn{1}{l}{\touchTwo}        & \multicolumn{1}{l}{4 (1)}       & \multicolumn{1}{l}{5 (0)}         & \multicolumn{1}{l}{6.4}            & \multicolumn{1}{l}{.05}                   \\ \hline \hline
\multicolumn{1}{l}{\multirow{2}{*}{\begin{tabular}[c]{@{}l@{}}Use for targets \\ in top region\end{tabular}}} & \multicolumn{1}{l}{\touchOne}        & \multicolumn{1}{l}{3 (2)}       & \multicolumn{1}{l}{4 (1)}         & \multicolumn{1}{l}{5.56}           & \multicolumn{1}{l}{.05}                   \\ 
\multicolumn{1}{l}{}                                                                                          & \multicolumn{1}{l}{\touchTwo}        & \multicolumn{1}{l}{4 (1)}       & \multicolumn{1}{l}{5 (1)}         & \multicolumn{1}{l}{6.23}           & \multicolumn{1}{l}{.05}                   \\ \hline
\multicolumn{1}{l}{\begin{tabular}[c]{@{}l@{}}Use for targets \\ in middle region\end{tabular}}               & \multicolumn{1}{l}{\touchTwo}        & \multicolumn{1}{l}{4 (.25)}    & \multicolumn{1}{l}{4 (1)}         & \multicolumn{1}{l}{5.33}           & \multicolumn{1}{l}{.05}                   \\ \hline
\multicolumn{1}{l}{\begin{tabular}[c]{@{}l@{}}Use for targets \\ in bottom region\end{tabular}}               & \multicolumn{1}{l}{\touchTwo}        & \multicolumn{1}{l}{4 (1)}       & \multicolumn{1}{l}{4 (1)}         & \multicolumn{1}{l}{11.27}          & \multicolumn{1}{l}{.001}                  \\ \hline
\multicolumn{1}{l}{Overall use}                                                                               & \multicolumn{1}{l}{\touchOne}        & \multicolumn{1}{l}{4 (1)}       & \multicolumn{1}{l}{4 (1)}         & \multicolumn{1}{l}{4.76}           & \multicolumn{1}{l}{.05}                   \\ \hline \hline
\end{tabular}%
}
\end{table*}

\begin{figure}[!t]
  \centering
  \includegraphics[width=\linewidth]{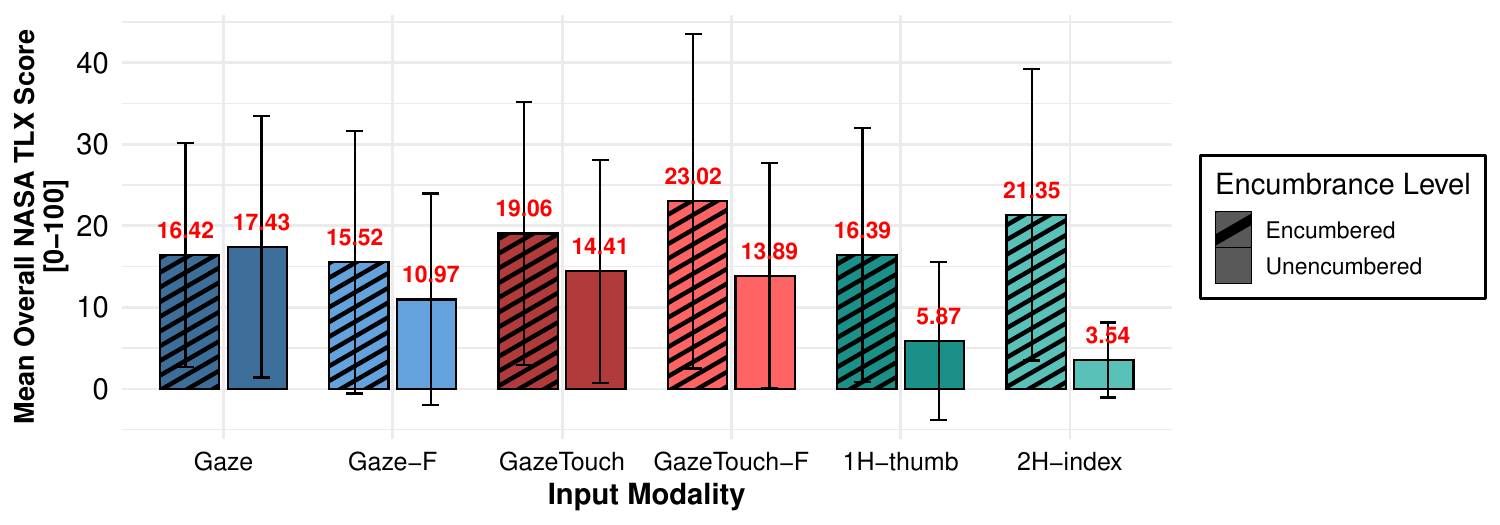}
    
  \caption{The mean overall NASA-TLX scores. The statistical tests showed that the perceived workload differed significantly due to encumbrance when using \gazetouchF, \touchOne, and \touchTwo, with participants experiencing a higher workload when encumbered than when unencumbered. The error bars represent the standard deviations.}
  \Description{The mean overall NASA-TLX scores. The statistical tests showed that the perceived workload differed significantly due to encumbrance when using \gazetouchF, \touchOne, and \touchTwo, with participants experiencing a higher workload when encumbered than when unencumbered. The error bars represent the standard deviations. }
  \label{fig:nasa_tlx}
\end{figure} 

\subsection{Perceived Cognitive Workload}
Figure \ref{fig:nasa_tlx} shows the mean overall NASA-TLX score averaged across the six dimensions. The scores are out of 100, with lower scores indicating a lower perceived cognitive workload. 

The statistical tests revealed a significant interaction between \textit{Input Modality} and \textit{Encumbrance Level} on the overall perceived workload (F\textsubscript{5,253} = 3.35, $p<.01$). 
We found no evidence that the perceived workload differed significantly across encumbrance levels for Gaze (F\textsubscript{1,23} = .36, $p>.05$) or \gazeF (F\textsubscript{1,23} = 4.20, $p>.05$).
While we found no significant differences in the perceived work due to encumbrance when using GazeTouch (F\textsubscript{1,23} = 4, $p>.05$), the differences were significant when using \gazetouchF (F\textsubscript{1,23} = 4.61, $p<.05$).  
\gazetouchF resulted in a significantly higher ($p<.05$) workload when participants were encumbered ($M=23.0$, $SD=20.5$, $p<.05$) than unencumbered ($M=13.9$, $SD=13.8$). 
The perceived workload differed significantly due to encumbrance when using \touchOne (F\textsubscript{1,23} = 26.45, $p<.0001$), and \touchTwo (F\textsubscript{1,23} = 67.21, $p>.0001$).
Encumbrance resulted in both \touchOne ($M=16.4$, $SD=15.6$) and \touchTwo ($M=21.4$, $SD=17.9$) to be perceived to cause significantly higher workload ($p<.0001$) compared to when used when unencumbered (\touchOne: $M=5.87$, $SD=9.68$, \touchTwo: $M=3.54$, $SD=4.59$).

\observation{\gazetouchF, and \touchOne and \touchTwo were perceived to cause significantly higher workload when participants were encumbered compared to when they were unencumbered. We found no significant differences in the perceived workload due to encumbrance on Gaze, \gazeF, and GazeTouch.}

\subsection{Perceived Usability}
\begin{figure*} [t!]
    \centering
    \begin{subfigure}[b]{.49\textwidth}
        \centering
        \includegraphics[width=\textwidth]{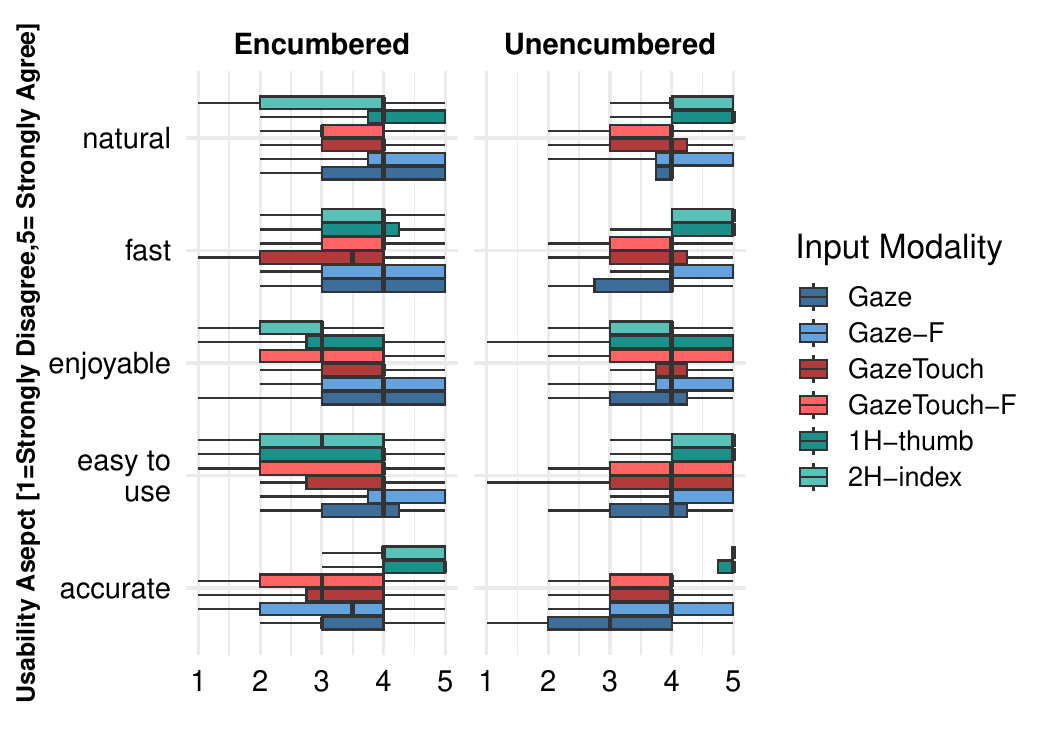}
        \caption{Usability aspects}
        \label{fig:likert_A}
    \end{subfigure} \hfill
    \begin{subfigure}[b]{.49\textwidth}
        \centering
        \includegraphics[width=\textwidth]{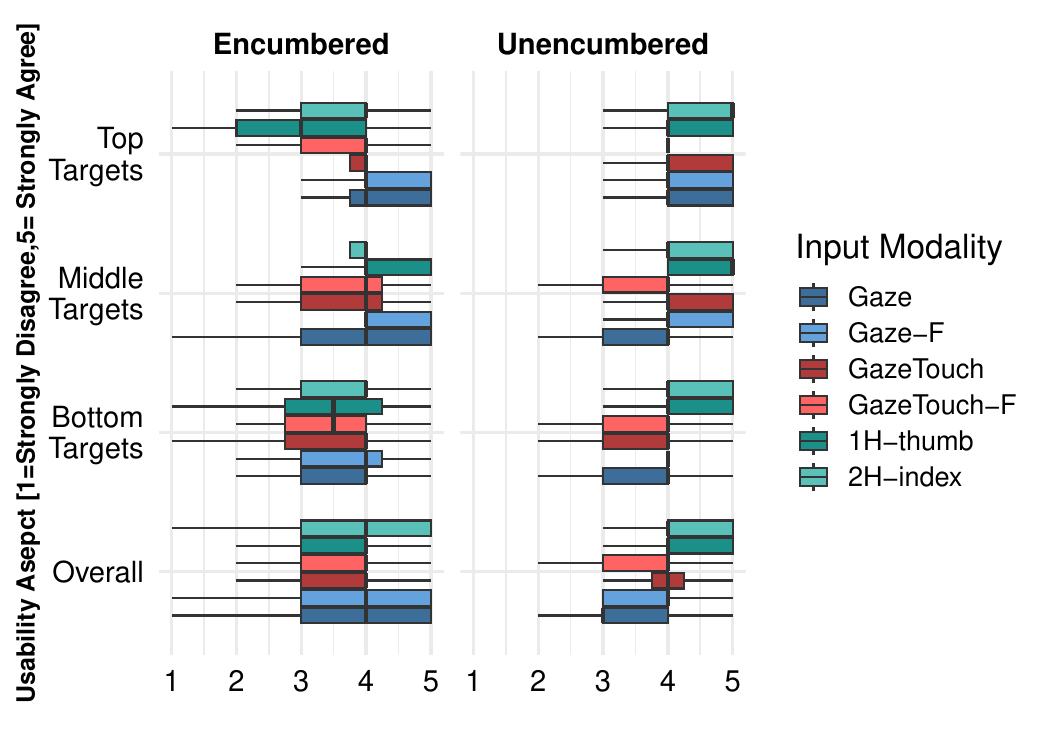}
        \caption{Participants’ willingness to use input modalities}
        \label{fig:likert_B}
    \end{subfigure}

    \caption{On a 5-point Likert scale (1= Strongly Disagree; 5= Strongly Agree), participants rated the usability of input modalities while walking encumbered and unencumbered. Boxes show the interquartile range, and the line inside each box represents the median.}
    \label{fig:likert_}
    \Description{}
\end{figure*}

Participants responded to Likert scale questions (1= Strongly Disagree; 5= Strongly Agree), to reflect on their experience with the six input modalities when used while encumbered and unencumbered. We evaluated five usability aspects using statements to determine whether participants found selection with each input modality to be natural, fast, enjoyable, easy to use, and accurate (see Figure \ref{fig:likert_A}). We further assessed participants' willingness to use each input modality across different screen regions (top, middle, bottom) and overall, regardless of region (see Figure \ref{fig:likert_B} and Section \ref{sec:participants_reflections}). We used the Friedman test for statistical significance, with the Wilcoxon signed-rank test for pairwise comparisons, corrected with the Bonferroni correction to account for multiple comparisons. Table \ref{tab_likert_significance} shows significant results from the Friedman tests. We report on general perceived usability across modalities, highlighting when results are statistically significant. The detailed findings are presented in Table \ref{tab_likert_significance_all}.

On average, across both encumbrance levels, participants were generally positive about Gaze and \gazeF, perceiving them as natural, fast, enjoyable, and easy to use. On the other hand, perceived accuracy received lower ratings than other aspects, particularly when participants were encumbered.

Both GazeTouch and \gazetouchF received positive ratings across all aspects, though ratings were slightly lower when walking encumbered, particularly for accuracy and speed. 
Perceiving \gazetouchF as natural was significantly impacted by encumbrance ($\chi^{2}(1)= 4.76$, $p<.05$). However, pairwise comparison showed that the differences were not significant ($p>.05$). 
Similarly, participants' perceptions of the accuracy varied significantly for GazeTouch ($\chi^{2}(1)= 5.44$, $p<.05$) and \gazetouchF ($\chi^{2}(1)= 5.4$, $p<.05$) across encumbrance levels. However, the pairwise comparisons did not show significant differences between pairs ($p>.05$).

\observation{Although participants’ perceptions of naturalness, speed, enjoyment, ease of use, and accuracy of each input modality varied somewhat across encumbrance levels, the differences observed among conditions Gaze, \gazeF, GazeTouch, and \gazetouchF were not statistically significant.}

Friedman tests showed significant differences in the participants' perception of how fast both \touchOne ($\chi^{2}(1)= 5.4$, $p<.05$) and \touchTwo ($\chi^{2}(1)= 13.24$, $p<.0005$) were for target selection due to encumbrance. Participants perceived \touchOne ($Mdn=4$, $IQR=1.25$) and \touchTwo ($Mdn=4$, $IQR=1$) as significantly slower when encumbered than when unencumbered (\touchOne: $Mdn=5$, $IQR=1$, $p<.5$; \touchTwo: $Mdn=5$, $IQR=1$, $p<.0005$). 
Regarding enjoyability, \touchOne ($\chi^{2}(1)= 8.05$, $p<.005$) and \touchTwo ($\chi^{2}(1)= 9.94$, $p<.005$) were perceived significantly differently depending on the levels of encumbrance. Participants perceived \touchOne ($Mdn=3$, $IQR=1.25$) and \touchTwo ($Mdn=3$, $IQR=1$) as significantly less enjoyable when they were encumbered than when unencumbered (\touchOne: $Mdn=4$, $IQR=2$, $p<.05$; \touchTwo: $Mdn=4$, $IQR=1$, $p<.005$).
The differences in the perceived ease of use were significant for \touchOne ($\chi^{2}(1)= 9$, $p<.005$) and \touchTwo ($\chi^{2}(1)= 14.22$, $p<.0005$). Participants perceived \touchOne ($Mdn=5$, $IQR=1$) and \touchTwo ($Mdn=5$, $IQR=1$) as significantly easier to use when unencumbered than when encumbered (\touchOne: $Mdn=4$, $IQR=2$, $p<.005$; \touchTwo: $Mdn=3$, $IQR=2$, $p<.0001$).
While both \touchOne and \touchTwo were perceived as more accurate when participants were unencumbered (\touchOne: $Mdn=5$, $IQR=.25$; \touchTwo: $Mdn=5$, $IQR=0$) compared to when they were encumbered (\touchOne: $Mdn=5$, $IQR=1$; \touchTwo: $Mdn=4$, $IQR=1$), the difference was significant for \touchTwo ($\chi^{2}(1)= 6.4$, $p<.05$). Participants perceived \touchTwo as significantly ($p<.05$) more accurate when unencumbered than when encumbered.

\observation{Participants perceived \touchOne and \touchTwo as significantly slower, less enjoyable, and harder to use, when they were encumbered compared to when they were unencumbered. \touchTwo was perceived as significantly less accurate when encumbered.}

\subsubsection{Participants' Willingness on Using the Input Modalities} \label{sec:participants_reflections}
While we found no evidence for significant differences in the participants' willingness due to encumbrance to use Gaze and GazeTouch, both with and without feedback for the entire screen or across the different regions, Friedman test revealed that participants' willingness to use both \touchOne and \touchTwo for top targets was significantly impacted by encumbrance levels (\touchOne: $\chi^{2}(1)=5.56$, $p<.05$; \touchTwo: $\chi^{2}(1)= 6.23$, $p<.05$). Participants were significantly ($p<.01$) less willing to use \touchOne when encumbered ($Mdn=3$, $IQR=2$) than when unencumbered ($Mdn=4$, $IQR=1$). They were also less willing ($p<.01$) to use \touchTwo for selection in the top region of the screen when encumbered ($Mdn=4$, $IQR=1$) than when unencumbered ($Mdn=5$, $IQR=1$).
Willingness to use \touchTwo for middle and bottom targets was also significantly impacted by encumbrance (Middle: $\chi^{2}(1)=5.33$, $p<.05$; Bottom: $\chi^{2}(1)=11.27$, $p<.001$). Pairwise comparisons showed that participants were significantly less willing to use \touchTwo for selection from the middle ($Mdn=4$, $IQR=.25$, $p<.05$) and bottom ($Mdn=4$, $IQR=1$, $p<.001$) regions when encumbered compared to when unencumbered. 
Regardless of the regions, the participants' willingness to use \touchOne for target selection was significantly impacted by the encumbrance levels ($\chi^{2}(1)=4.76$, $p<.05$). Participants were significantly ($p<.05$) more willing to use \touchOne for selection when unencumbered ($Mdn=4$, $IQR=1$) compared to when encumbered ($Mdn=4$, $IQR=1$).

\observation{Participants were significantly less willing to use \touchOne and \touchTwo to select targets from the top regions of the screen when encumbered, compared to when unencumbered. Willingness to use \touchTwo to select from the middle and bottom regions was also significantly affected by encumbrance. Overall, participants were significantly less willing to use \touchOne when encumbered than when unencumbered.}

\subsubsection{Comparing Perceived Usability Across Input Modalities}
% natural no
We conducted statistical tests to determine whether differences across input modalities were significant for each perceived usability aspect, for each encumbrance level. We only reported results showing statistically significant differences.

\textit{When walking encumbered}, the Friedman test showed significant differences between the input modalities in how natural ($\chi\textsuperscript{2}(5)$ = 18.76, $p<.005$), enjoyable ($\chi\textsuperscript{2}(5)$ = 23.59, $p<.0005$), easy to use ($\chi\textsuperscript{2}(5)$ = 13.04, $p<.05$), and accurate ($\chi\textsuperscript{2}(5)$ = 49.78, $p<.0001$) the input modalities were perceived. 
Post hoc analysis showed that \touchTwo ($Mdn=3$, $IQR=1$, $p<.01$) was perceived as significantly ($p<.05$) less enjoyable compared to Gaze ($Mdn=4$, $IQR=2$) and \gazeF ($Mdn=4$, $IQR=2$). 
\touchOne and \touchTwo were perceived as significantly more accurate than Gaze and GazeTouch with and without feedback. 

\textit{When walking unencumbered}, the Friedman test showed significant differences between the input modalities in how natural ($\chi\textsuperscript{2}(5)$ = 14.64, $p<.05$), fast ($\chi\textsuperscript{2}(5)$ = 34.99, $p<.0005$), easy to use ($\chi\textsuperscript{2}(5)$ = 22.48, $p<.0005$), and accurate ($\chi\textsuperscript{2}(5)$ = 57.14, $p<.0001$) the input modalities were perceived. 
Post hoc analysis showed that \touchTwo ($Mdn=5$, $IQR=1$) was perceived as significantly faster than Gaze ($Mdn=4$, $IQR=1.25$, $p<.01$), GazeTouch ($Mdn=4$, $IQR=1.25$, $p<.01$), \gazetouchF ($Mdn=4$, $IQR=1$, $p<.005$). \touchTwo ($Mdn=5$, $IQR=1$) was perceived as significantly easier ($p<.05$) to use compared to Gaze ($Mdn=4$, $IQR=1.25$).
\touchOne and \touchTwo were perceived as significantly more accurate than Gaze and GazeTouch with and without feedback. 
We found significant differences between input modalities in participants' willingness to use them across the screen ($\chi\textsuperscript{2}(5)$ = 17.95, $p<.005$), and for the middle ($\chi\textsuperscript{2}(5)$ = 28.75, $p<.0005$) and bottom regions of the screen ($\chi\textsuperscript{2}(5)$ = 20.96, $p<.001$).

\begin{figure}[!t]
  \centering
  \includegraphics[width=\linewidth]{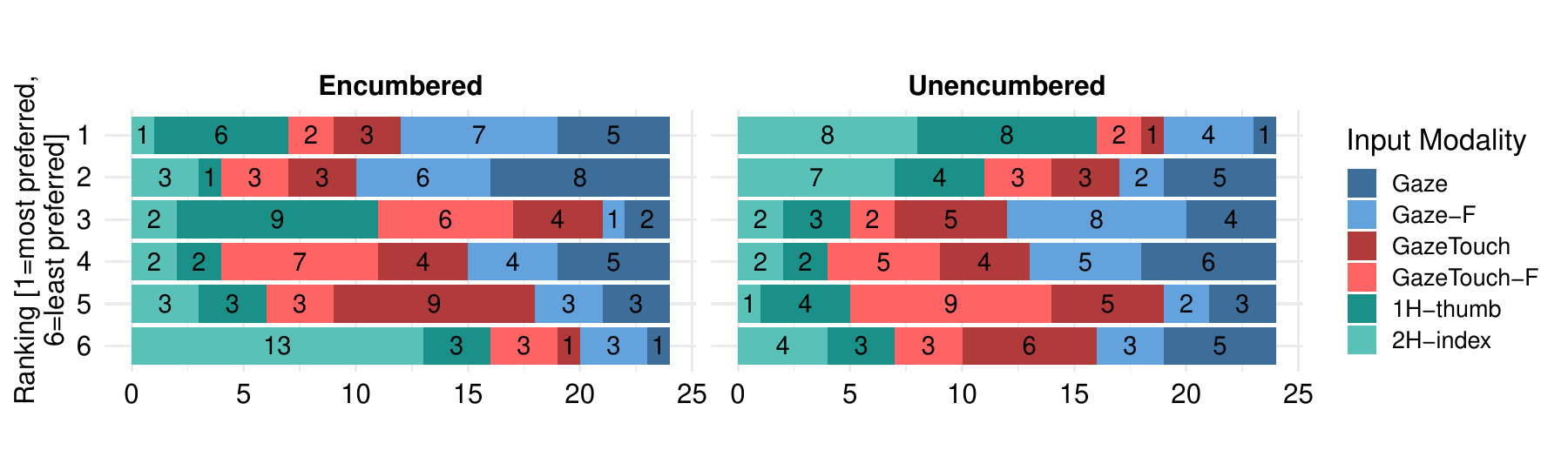}
    
  \caption{Participants reported their preference for the input modalities in each encumberance level, where 1 is the most preferred, to 6 as the least preferred. While Gaze and \gazeF were the most preferred input modality while walking encumbered \textbf{(Left)}, \touchOne and \touchTwo were the most preferred when walking unencumbered  \textbf{(Right)}. The numbers represent the counts for each input modality in each ranking.}
  \Description{}
  \label{fig:ranking}
\end{figure}

\begin{table*}[]
\centering
\caption{Participants reported their preference for each input modality based on encumbrance levels. Input modalities are ordered from most preferred to least preferred. While Gaze and \gazeF were the most preferred for target selection while walking encumbered, \touchTwo and \touchOne were the most preferred when participants were unencumbered.}
\label{tab:ranking}
\resizebox{.9\linewidth}{!}{%
\begin{tabular}{llllllll}

\multicolumn{3}{c}{\textbf{Encumbered}}                                                                       &           &           & \multicolumn{3}{c}{\textbf{Unencumbered}}                                                                     \\ \cline{1-3} \cline{6-8} 
\multicolumn{1}{l|}{\textbf{Ranking}} & \multicolumn{1}{l|}{\textbf{Input Modality}} & \textbf{Weighted Score} & \textbf{} & \textbf{} & \multicolumn{1}{l|}{\textbf{Ranking}} & \multicolumn{1}{l|}{\textbf{Input Modality}} & \textbf{Weighted Score} \\ \cline{1-3} \cline{6-8} 
\multicolumn{1}{l|}{1}                & \multicolumn{1}{l|}{Gaze}                    & 100                     &           &           & \multicolumn{1}{l|}{1}                & \multicolumn{1}{l|}{\touchTwo}        & 103                     \\ \cline{1-3} \cline{6-8} 
\multicolumn{1}{l|}{2}                & \multicolumn{1}{l|}{\gazeF}         & 97                      &           &           & \multicolumn{1}{l|}{2}                & \multicolumn{1}{l|}{\touchOne}        & 97                      \\ \cline{1-3} \cline{6-8} 
\multicolumn{1}{l|}{3}                & \multicolumn{1}{l|}{\touchOne}        & 92                      &           &           & \multicolumn{1}{l|}{3}                & \multicolumn{1}{l|}{\gazeF}         & 88                      \\ \cline{1-3} \cline{6-8} 
\multicolumn{1}{l|}{4}                & \multicolumn{1}{l|}{\gazetouchF}    & 81                      &           &           & \multicolumn{1}{l|}{4}                & \multicolumn{1}{l|}{Gaze}                    & 76                      \\ \cline{1-3} \cline{6-8} 
\multicolumn{1}{l|}{5}                & \multicolumn{1}{l|}{GazeTouch}               & 80                      &           &           & \multicolumn{1}{l|}{5}                & \multicolumn{1}{l|}{\gazetouchF}    & 71                      \\ \cline{1-3} \cline{6-8} 
\multicolumn{1}{l|}{6}                & \multicolumn{1}{l|}{\touchTwo}        & 54                      &           &           & \multicolumn{1}{l|}{6}                & \multicolumn{1}{l|}{GazeTouch}               & 69                      \\ \cline{1-3} \cline{6-8} 
\end{tabular}}%
\end{table*}

\subsection{Input Modalities Ranking}
To explore the impact of encumbrance on preference for input modalities, after completing each block in the experiment based on the \textit{encumbrance levels}, we asked the participants to rank their preference for each input modality (1= Most preferred; 6= Least preferred). To compute the overall ranking score for each modality, we replaced raw scores with their weight factors to reflect the importance of each rank \cite{10.1145/3544548.3580871}.
Each rank was assigned a corresponding weight, with rank 1 receiving a weight of 6, decreasing by one point for each subsequent rank, with rank 6 receiving a weight of 1. These weighted ranks were summed to calculate the total weighted score for each input modality. Figure \ref{fig:ranking} and Table \ref{tab:ranking} show the counts for each input modality in each rank based on encumbrance level, and also the total weighted scores, ordered by the most preferred input modality to the least preferred based on encumbrance level. We conducted statistical tests using Friedman's test, with pairwise comparisons using the Wilcoxon signed-rank test to examine the impact of encumbrance on preferences for each input modality. While participants preferred Gaze the most ($score=100$), with 5 ranking it first, and 8 as second, followed by \gazeF ($score=97$) when they were encumbered, both modalities were ranked third and fourth when participants were unencumbered (Gaze: $score=76$, \gazeF: $score=88$). There was significant differences in the preference for Gaze based on encumbrance level ($\chi\textsuperscript{2}(1)$ = 5.56, $p<.05$). Gaze was significantly ($p<.05$) more preferred when participants were encumbered ($Mdn=2$, $IQR=2$) compared to when they were not ($Mdn=4$, $IQR=2.25$). We found no significant difference in the preference for \gazeF depending on the encumbrance level ($\chi\textsuperscript{2}(1)$ = .47, $p>.05$). 
GazeTouch did not seem to be favoured by participants in both levels of encumbrance, being ranked the second least favoured when encumbered ($score=80$), and the least favoured when unencumbered ($score=69$). Although \gazetouchF was the second least favoured when unencumbered ($score=71$), it was ranked as the fourth most favoured when encumbered ($score=81$). The statistical tests did not reveal significant differences depending on the encumbrance on the preference of GazeTouch ($\chi\textsuperscript{2}(1)$ = 3.77, $p>.05$), and \gazetouchF ($\chi\textsuperscript{2}(1)$ = .89, $p>.05$). 
\touchTwo was ranked as the most preferred input modality for target selection when participants were walking unencumbered ($score=103$), with 8 ranking it first and 7 as second, followed by the \touchOne ($score=97$). However, both input modalities were ranked third (\touchOne: $score=92$) and sixth (\touchTwo: $score=54$) as the least favoured when participants were encumbered.
While we found no significant difference in the preference for \touchOne depending on the encumbrance ($\chi\textsuperscript{2}(1)$ = 0, $p>.05$), there was a significant difference in the preference for \touchTwo based on the encumbrance level ($\chi\textsuperscript{2}(1)$ = 18, $p<.0005$). Participants significantly ($p<.001$) ranked \touchTwo as less preferred when encumbered ($Mdn=6$, $IQR=2.25$) compared to when they were not ($Mdn=2$, $IQR=3$).

\observation{When encumbered, while Gaze and \gazeF were ranked as the most preferred input modalities, Gaze was significantly ranked better when participants were encumbered than unencumbered. When unencumbered, while \touchTwo and \touchOne were ranked as the most preferred input modalities, \touchTwo was significantly ranked better when participants were unencumbered than when they were encumbered.}

\subsection{Qualitative Feedback}
We interviewed participants at the end of the experiment to reflect on their experience with using various input modalities while walking, and being encumbered. We employed inductive thematic analysis to analyse the interview data. The leading researcher reviewed recordings and corrected errors in the auto-generated transcripts, coded the data using QCamps \cite{qcamap}, developed categories, and generated final themes.

\subsubsection{How Encumbrance Affects Gaze During Walking}
Out of the $24$ participants, $17$ reported positive experiences with Gaze when walking encumbered. They reflected on aspects such as ease of use ($n=10$), convenience ($n=4$), avoidance of touch due to hand constraints ($n=4$), reduced cognitive and physical effort ($n=3$), and balance mental and physical demand ($n=2$), \textit{\say{Without weight, touch is convenient, with the weights, gaze is convenient}}(P8). 
Regardless of the encumbrance level, negative aspects in using Gaze, were mentioned by 12 participants, such as inaccuracy ($n=3$), slow ($n=3$), personal safety ($n=2$).

\subsubsection{How Encumbrance Affects GazeTouch During Walking}
Out of the $24$ participants, $18$ reported positive aspects of using GazeTouch for selection, such as generally liking the technique ($n=7$). Eight participants reported positive experiences using GazeTouch while encumbered, describing it as convenient and easy to use ($n=4$) and noting that touching the screen to confirm selection made them feel more in control ($n=4$). Some appreciated that they can tap anywhere to select ($n=3$).
On the other hand, while $14$ participants mentioned negative aspects of using GazeTouch, reporting that it causes cognitive and physical effort ($n=3$), five reflected on their negative experience, particularly when encumbered, describing it as demanding mainly because it involves touch ($n=4$).
Four participants reported that GazeTouch requires getting used to it, but it becomes easier and more natural with practice. \textit{\say{First, I found [GazeTouch] not natural for me, but then, when I got used to it, especially when I carry bags, it is very convenient for me.}} (P3).

\subsubsection{How Encumbrance Affects \touchOne and \touchTwo During Walking}
Reporting generally on touch input, 10 participants mentioned negative experiences when using touch while encumbered, mentioning physical effort or discomfort ($n=4$), inconvenience or frustration ($n=6$), cognitive load or distraction ($n=2$), performance or control issues ($n=2$), and preference to avoid touch ($n=2$). P17 described their experience when they were encumbered, \textit{\say{I feel like when you are encumbered, it definitely makes the physical challenge of using your hand, whether it is one-handed or \touchTwo, much much more difficult}}.
When using \touchOne while encumbered, $13$ of $24$ participants reported negative experiences, describing it as physically demanding or difficult ($n=8$) and inconvenient or uncomfortable ($n=2$). Some were also afraid of dropping the phone ($n=4$), while four reported that screen size was an issue. P7 reported that, \textit{\say{when you carry weights in your hands, it is a lot of effort [to use touch]}}, and P18, \textit{\say{if the task had been harder or the bags had been heavier, touch becomes you can not really do it while walking}}. 
Participants reported positive experiences using \touchOne regardless of the encumbrance level ($n=12$), because they are used to it ($n=4$), or because they found \touchOne fast ($n=2$). P9 reported on ranking touch higher than Gaze, \textit{\say{because that is just what I am used to, and that is what comes naturally}}. Five participants were positive about using \touchOne when encumbered, describing it as natural ($n=4$), while one noted that it was less annoying and did not require positioning the phone in the middle of the body to carry bags and touch with the index finger, unlike \touchTwo. P13 noted that they preferred \touchOne reporting that, \textit{\say{I feel it is natural for me to use [one hand touch], with the \touchTwo [touch], it is not natural for me to use my index.}}, \textit{\say{It is just that [touch] one-handed just comes naturally, so I think it is a bit faster, because you are just used to reacting with your thumb}} (P9).
For \touchTwo, $10$ participants reported negative experiences when using it while being encumbered, describing it as physically demanding and effortful ($n=4$), inconvenient and difficult to perform while holding bags ($n=5$), and inaccurate ($n=2$). 
P17 reported on \touchTwo \textit{\say{it was very tricky [when encumbered], and also I could not get a decent touch on it, so I had to touch the target multiple times, whereas I think without the bags, it was very easy.}} On the other hand, only one participant mentioned that it was easy to use \touchTwo when they were encumbered.

\subsubsection{Effects of Visual Feedback on Gaze-Based Input During Walking}
Overall, 13 participants reported positive experiences with gaze-based input modalities that provided visual feedback: seven with Gaze and five with GazeTouch. When using Gaze, two participants reported a preference for visual feedback to remove any confusion about whether the phone is tracking them, to adjust their gaze ($n=1$), or to ensure the phone is tracking them accurately ($n=1$). 
When using GazeTouch without visual feedback, P20 mentioned that \textit{\say{I had to touch it repeatedly, and I was confused why it is not accepting [my input]}}, and P4 reported that, \textit{\say{I was just tapping until it magically landed up}}, and P19 reported that \textit{\say{I feel the false input rate would be quite high}}.
On the other hand, 14 participants reported a negative experience with visual feedback: 10 when using Gaze and six when using GazeTouch. Participants found the visual feedback when using Gaze distracting ($n=3$), slower ($n=2$), and diverted attention when using GazeTouch ($n=2$).

\subsubsection{Input Modality Reflections Across Target Regions}
Some participants reported positive perceptions of using Gaze ($n=8$), Gazetouch ($n=2$), and \touchTwo ($n=1$) to select targets from the top regions of the mobile screen. While some participants reported difficulties using Gaze ($n=6$) and \touchOne ($n=4$) in the bottom region, two participants found Gaze easy to use, and four found \touchOne easy or preferred for the bottom region. Six participants reported that with \touchOne, areas farther away or on the opposite side were harder to reach with the thumb, making corner and opposite-side selections more difficult, as P9 noted that \textit{\say{I tend to use my phone one handed, even though it does have its limitations, like you saw there was the one in the bottom right corner that I was instinctively wanting to touch with my other hand}}, and P10, \textit{\say{\touchOne is obviously a little bit problematic on the corners, which is the non dominant corner, so the thumb cannot reach}}.

\section{Discussion \& Future work}
In this section, we discuss the detailed effects of encumbrance on gaze- and touch-based input modalities for target selection on handheld mobile devices. 

\subsection{Effect of Encumbrance on Gaze- and Touch-based Input}

Our study confirms prior work on the impact of encumbrance on touch interaction, while offering novel insights into its effects on gaze.

\subsubsection{Touch selection is significantly impacted by encumbrance}
In our study, while the results showed that selecting targets with touch inputs was generally faster than selecting with gaze-based inputs, which was expected, our findings demonstrated that \touchOne and \touchTwo selections were significantly affected by encumbrance, where participants took significantly longer to select when encumbered than when unencumbered. This is consistent with previous research, which found that selection time was significantly longer when participants performed selection with touch using three different input postures \cite{10.1145/2556288.2557312}, although the target sizes used in our experiment were considered large for touch interfaces. 
The findings also aligned with participants' perceptions of touch-based inputs, which they perceived as significantly slower when performing the tasks while encumbered. 
For task completion rate, our results using ANOVA showed that participants completed significantly less tasks when using \touchOne while walking encumbered.
% The analysis revealed substantial evidence that encumbrance affects task completion rate when using \touchTwo, where less tasks were completed when encumbered.
Our findings align with work showing that, for touch input, participants were significantly more accurate when unencumbered than when carrying bags, across three input postures (two-handed index finger, one-handed preferred thumb, and two-handed both thumbs) for target selection on mobile devices \cite{10.1145/2556288.2557312}. 
While the ANOVA results aligned with prior work for \touchOne, the misalignment observed with two-handed index-finger input might be due to differences in target size between our experiment and prior work, as well as differences in touch technology. 
That said, the Bayesian analysis provided substantial evidence for differences in task completion rates (referred to as accuracy in prior work \cite{10.1145/2556288.2557312}), suggesting that \touchTwo is impacted by encumbrance.
For error counts, the Bayesian analysis revealed that encumbrance affected both \touchOne and \touchTwo, with substantial evidence for \touchOne and very strong evidence for \touchTwo, suggesting that walking while encumbered increases the likelihood of accidentally selecting incorrect targets.

\subsubsection{Gaze with Visual Feedback maintained stable performance under encumbrance}
For selection time, we found no significant differences in selection time due to encumbrance when using Gaze, \gazeF, GazeTouch, or \gazetouchF.
Bayesian analysis also supported this, with no evidence of strong encumbrance effects, suggesting comparable performance when selecting targets under encumbrance and unencumbrance. 
For task completion rate, our results using ANOVA showed that participants completed significantly less tasks when using GazeTouch and \gazetouchF walking encumbered. 
On the contrary, the task completion rates for Gaze and \gazeF when walking encumbered and unencumbered were comparable, with anecdotal or no evidence for differences due to encumbrance, respectively, as revealed by the Bayesian analysis. 
For error counts, the Bayesian analysis revealed that encumbrance affected Gaze with strong evidence, whereas no such effect was observed when using \gazeF. This highlights the benefit of visual feedback in maintaining consistent error counts.
GazeTouch and \gazetouchF also resulted in comparable error counts when used for selection, while being encumbered and unencumbered.
These findings suggest that gaze-based input techniques, especially Gaze with visual feedback (\gazeF), were more robust to performance changes due to encumbrance.
Although not significant, the unexpected increase in errors with GazeTouch when participants were unencumbered, compared to when they were encumbered, could be attributed to several factors, as supported by qualitative data and experimenter observations. 
First, the novelty of the input technique (GazeTouch) to participants. Additionally, in the absence of feedback, some participants repeatedly tapped the interface as soon as targets appeared, which may have contributed to higher error counts in some trials. We expect that, because tapping was easier when participants were unencumbered, as many suggested, they tapped more to select faster (e.g., P4, P19, P20), which probably led to higher error counts when walking unencumbered.

\subsection{Users' Expectation may Lead to Frustration}
Our findings showed that touch input is more affected than gaze-based input, as selection time and completion rate varied. This was also reflected in participants' Likert-scale responses regarding the usability of touch input, in which they found \touchOne and \touchTwo significantly slower, less enjoyable, and more difficult to use when encumbered. 
They also perceived \touchTwo as significantly less accurate when encumbered. 
Participants also perceived \touchOne and \touchTwo to cause significantly higher workload when used while encumbered. They described touch input as inconvenient or frustrating, or as requiring physical effort or causing discomfort. 
Such a negative experience with touch while encumbered may stem from participants' expectations of consistent performance with using touch. This variation in performance led participants to experience inconvenience and discomfort. Prior work shows that when technology falls short of user expectations, it leads to user frustration \cite{Ferreri2021, Taylor2022}. It is important to consider the negative influence of technology on users, as such frustrating experiences may impact a person's emotional well-being \cite{Taylor2022}. 
Depending on the context in which touch performance is likely to change or degrade, such as when users are encumbered, designers of user interfaces should consider this impact on the user experience. When the context warrants, they may offer options that align with users' preferences, such as switching to another modality, recommending and enabling another modality (e.g., Gaze), and letting users decide what to use, based on their preferences \cite{LaViola2015Multimodal}. Future work could investigate the effect of modality switching depending on context, such as between touch and Gaze when users are encumbered, on usability. 
This includes exploring automatic switching between modalities or displaying an option to switch to another one when encumbrance is detected.
Building on prior work that used accelerometers and gyroscope sensors attached to the feet to classify differences between walking normally, walking while carrying a glass of water, and walking blindfolded \cite{6252501}, inertial sensor data from phones, an area that warrants further exploration, can be leveraged to detect encumbrance.

\subsection{Target Regions and \touchOne}
Our findings showed that \touchOne was generally affected by the target region, with selection performance varying across regions. Top targets were generally slower to select than targets in the middle and bottom regions, whereas selecting targets from the middle region was the fastest overall. 
Even in the subjective rating, participants showed less willingness to use \touchOne when selecting targets in the top region of the screen. The fast performance in the middle region could be due to how participants held the mobile device during the experiment, as prior work has shown that the comfortable area for holding the device with the thumb is in the middle of the screen \cite{10.1145/3173574.3173605}. Such an area does not require changing grip or losing stability by overstretching, which could result in faster selection.
% Prior work showed that the upper half of the smartphone is  
% Participants showed less willingness to use a \touchOne when encumbered than when unencumbered. 
Prior work has shown that targets near edges and corners \cite{10.1145/3279778.3279791} and those farther from the thumb \cite{10.1145/2556288.2557312} are difficult to select. Huy Veit Le et al.~showed that the lower left quarter targets to right-handed users were not reachable by any finger without a grip change \cite{10.1145/3173574.3173605}. 
Selection from such areas using the thumb requires flexing or stretching the thumb \cite{10.1145/3279778.3279791}, or changing grip \cite{10.1145/3173574.3173605}. 
In our experiment, some participants reported using a coping mechanism of adjusting the grip to reach the corner targets or to counteract the swinging bags in the hand. However, this could lead to dropping the phone \cite{10.1145/3173574.3173605}. The problem exacerbates when carrying bags, as some participants reported fearing that they would drop the device.  
Given that users prefer to use phones one-handed \cite{Ng2013Encumbrance, Karlson2006, 10.1145/3173574.3173605}, with the influence of hand size and screen size of mobile devices on touch input \cite{10.1145/3279778.3279791}, when users are encumbered, gaze can serve as a potential alternative modality to support them in such situations, mitigating grip changes. 
While some participants experienced issues using Gaze in the bottom region of the screen, an observation aligned with prior research \cite{10.1145/3706598.3713092}, advancements in eye-tracking technology may make this limitation negligible. 
Previous work suggested increasing target heights in these regions to enhance accuracy and mitigate this limitation \cite{10.1145/3706598.3713092, 10.1145/3025453.3025599}. We recommend exploring other eye movements that do not rely on the absolute point of gaze, such as Pursuits \cite{10.1145/2493432.2493477} and gestures \cite{10.1145/1743666.1743710}, which may help mitigate such issues in these regions.

\subsection{Enhancing Gaze Performance Consistency with Visual Feedback}
The results suggested that continuous feedback on in-focus targets before selection improved the consistency of Gaze input, with participants' performance comparable whether or not bags were carried. However, subjective experience revealed mixed opinions about using Gaze and GazeTouch with or without feedback. Some participants described the feedback as distracting and diverting their attention; a similar experience was observed among participants in prior work when using feedback for eye typing on a desktop machine, where they found the feedback confusing even though their performance was reasonably good \cite{10.1145/968363.968390}. However, some other participants appreciated the visual feedback for helping them refine their selection, reassuring themselves that their eyes were being tracked, and reducing the false input rate. 
We believe that further refinement of the visual feedback, in terms of colours and presentation, together with adjustable dwell time \cite{Majaranta2006}, can enhance the user experience and avoid distractions. The visual feedback can be carefully designed as an on-target animated indicator that gradually fills continuously to indicate the time remaining for selection. Such a design can be borrowed from prior work \cite{10.1145/3419249.3420122}. For GazeTouch, we recommend showing visual feedback only when gaze is steady on a target to avoid distraction, and hiding it when eyes are moving.

\subsection{Impact of Encumbrance on Subjective Measures}
While touch input generally outperformed gaze-based input in selection time, completion rate, and error counts, participants mostly favoured gaze input when walking while encumbered. The differences in perceived usability between gaze- and touch-based inputs were insignificant in many aspects, as reflected in the participants' perceptions of how fast, natural, and easy to use the input modalities were (see Figure \ref{fig:likert_} and Table \ref{tab_likert_significance}). While this could be due to novelty effects associated with gaze input, several participants who prioritised touch did so because of its familiarity, as reported by P9 and a few other participants. The general lack of preference for touch inputs may stem from encumbrance. Many participants reported that when carrying bags, they preferred using Gaze, whereas when unencumbered, they preferred touch.
Such misalignment between objective performance and subjective preferences is not uncommon in HCI. For example, when Emanuel et al.~compared the usability of Pattern- and PIN-based authentication on mobile devices \cite{10.1145/2493190.2493231}, although users authenticated significantly faster using PIN and made more errors using patterns, they rated pattern-based authentication as fast, easy to use, and faster at recovering from errors. They attributed this contradiction to users' liking of pattern authentication. Similarly, in our experiment, the preference for gaze over touch inputs when participants were encumbered could be attributed to their enjoyment of gaze input, as reflected in their ratings of the enjoyment of the input modalities during selection while encumbered. These findings suggest that we should not exclude gaze solely based on objective performance, as there may be situations and contexts where gaze is favoured. Potential future improvements may result from advances in computer vision to enhance eye tracking on handheld mobile devices, thereby improving accuracy and widening the scenarios in which gaze input can be effectively employed.

\subsection{Not Captured by Quantitative Data}
Results from qualitative data showed that a few participants feared dropping the mobile device while walking encumbered, and using \touchOne. Such an incident is likely to happen due to the current smartphone's screen size, which requires changing grip to reach near edges and corner targets \cite{10.1145/3279778.3279791}, and encumbrance increases this possibility. 
Such fear was evident in prior work where patients diagnosed with Parkinson's disease mentioned difficulties holding their phone, and feared dropping the mobile device \cite{10.1145/3338286.3340133}. In another study, researchers examined older adults' experience with wearing sensors to detect falls. Older adults reported fear of dropping or damaging wearable sensors \cite{Tornblom2025}. The fear of breaking technology \cite{Wilson2023}, being without a mobile phone (NOMOPHOBIA) \cite{Bhattacharya2019}, or of health-related issues \cite{Vaportzis2017} could be a barrier to using such technologies, especially in a mobile context. When using Gaze while being encumbered, this burden can potentially be alleviated by allowing users to select targets with their gaze while maintaining a firm grip on the device, eliminating the need to change grips or risk dropping the phone during interaction. 
On the other hand, some participants expressed concerns about using gaze as an input modality. They felt that personal safety might be affected if they had to focus on the mobile device more than necessary while walking, for example. Developing mechanisms that preserve the user's state while they are walking, so they do not have to re-enter input from scratch, could be a potential direction \cite{10.1145/3544548.3580871, 10.1145/3419249.3420122}. This would help reassure users that they can maintain awareness of their surroundings while still providing input without losing their progress.

\section{Conclusion}
In this work, we explored the effect of encumbrance on the performance of three input modalities: 1) Gaze using Dwell time (with/without visual feedback), 2) GazeTouch (with/without visual feedback), and 3) one- or two-handed touch. 
We found that gaze-based inputs, especially Gaze when used with visual feedback (\gazeF), are generally less affected by encumbrance than touch-based inputs, resulting in consistent performance. \gazetouchF, and \touchOne and \touchTwo were perceived to cause significantly higher workload when used while being encumbered. Participants preferred gaze most when encumbered, while they preferred touch most when unencumbered. We concluded with a discussion to enhance our understanding of the effects of encumbrance on gaze- and touch-based input modalities, and to shift toward selecting appropriate input modalities based on users' context when interacting with handheld mobile devices.

\bibliographystyle{ACM-Reference-Format}
\bibliography{sample-base}

@String{Computing = "Computing" }

@String{Computer = "{IEEE} Computer" }

@String{Springer = "Springer-Verlag" }

@inproceedings{10.1145/3025453.3025599,
author = {Feit, Anna Maria and Williams, Shane and Toledo, Arturo and Paradiso, Ann and Kulkarni, Harish and Kane, Shaun and Morris, Meredith Ringel},
title = {Toward Everyday Gaze Input: Accuracy and Precision of Eye Tracking and Implications for Design},
year = {2017},
isbn = {9781450346559},
publisher = {Association for Computing Machinery},
address = {New York, NY, USA},
url = {https://doi.org/10.1145/3025453.3025599},
doi = {10.1145/3025453.3025599},
booktitle = {Proceedings of the 2017 CHI Conference on Human Factors in Computing Systems},
pages = {1118–1130},
numpages = {13},
location = {Denver, Colorado, USA},
series = {CHI '17}
}

@inproceedings{10.1145/3229434.3229452,
author = {Khamis, Mohamed and Alt, Florian and Bulling, Andreas},
title = {The Past, Present, and Future of Gaze-Enabled Handheld Mobile Devices: Survey and Lessons Learned},
year = {2018},
isbn = {9781450358989},
publisher = {Association for Computing Machinery},
address = {New York, NY, USA},
url = {https://doi.org/10.1145/3229434.3229452},
doi = {10.1145/3229434.3229452},
booktitle = {Proceedings of the 20th International Conference on Human-Computer Interaction with Mobile Devices and Services},
articleno = {38},
numpages = {17},
location = {Barcelona, Spain},
series = {MobileHCI '18}
}

@inproceedings{10.1145/3025453.3025794,
author = {Huang, Michael Xuelin and Li, Jiajia and Ngai, Grace and Leong, Hong Va},
title = {ScreenGlint: Practical, In-Situ Gaze Estimation on Smartphones},
year = {2017},
isbn = {9781450346559},
publisher = {Association for Computing Machinery},
address = {New York, NY, USA},
url = {https://doi.org/10.1145/3025453.3025794},
doi = {10.1145/3025453.3025794},
booktitle = {Proceedings of the 2017 CHI Conference on Human Factors in Computing Systems},
pages = {2546–2557},
numpages = {12},
location = {Denver, Colorado, USA},
series = {CHI '17}
}

@inproceedings{10.1145/2168556.2168601,
author = {Dybdal, Morten Lund and Agustin, Javier San and Hansen, John Paulin},
title = {Gaze Input for Mobile Devices by Dwell and Gestures},
year = {2012},
isbn = {9781450312219},
publisher = {Association for Computing Machinery},
address = {New York, NY, USA},
url = {https://doi.org/10.1145/2168556.2168601},
doi = {10.1145/2168556.2168601},
booktitle = {Proceedings of the Symposium on Eye Tracking Research and Applications},
pages = {225–228},
numpages = {4},
location = {Santa Barbara, California},
series = {ETRA '12}
}

@inproceedings{10.5555/1778331.1778385,
author = {Drewes, Heiko and Schmidt, Albrecht},
title = {Interacting with the Computer Using Gaze Gestures},
year = {2007},
isbn = {3540747990},
publisher = {Springer-Verlag},
address = {Berlin, Heidelberg},
url = {https://doi.org/10.1007/978-3-540-74800-7_43},
doi = {10.1007/978-3-540-74800-7_43},
booktitle = {Proceedings of the 11th IFIP TC 13 International Conference on Human-Computer Interaction - Volume Part II},
pages = {475–488},
numpages = {14},
location = {Rio de Janeiro, Brazil},
series = {INTERACT'07}
}

@inproceedings{10.1145/2493432.2493477,
author = {Vidal, M\'{e}lodie and Bulling, Andreas and Gellersen, Hans},
title = {Pursuits: Spontaneous Interaction with Displays Based on Smooth Pursuit Eye Movement and Moving Targets},
year = {2013},
isbn = {9781450317702},
publisher = {Association for Computing Machinery},
address = {New York, NY, USA},
url = {https://doi.org/10.1145/2493432.2493477},
doi = {10.1145/2493432.2493477},
booktitle = {Proceedings of the 2013 ACM International Joint Conference on Pervasive and Ubiquitous Computing},
pages = {439–448},
numpages = {10},
location = {Zurich, Switzerland},
series = {UbiComp '13}
}

@inproceedings{10.1145/2556288.2557040,
author = {Kangas, Jari and Akkil, Deepak and Rantala, Jussi and Isokoski, Poika and Majaranta, P\"{a}ivi and Raisamo, Roope},
title = {Gaze Gestures and Haptic Feedback in Mobile Devices},
year = {2014},
isbn = {9781450324731},
publisher = {Association for Computing Machinery},
address = {New York, NY, USA},
url = {https://doi.org/10.1145/2556288.2557040},
doi = {10.1145/2556288.2557040},
booktitle = {Proceedings of the SIGCHI Conference on Human Factors in Computing Systems},
pages = {435–438},
numpages = {4},
location = {Toronto, Ontario, Canada},
series = {CHI '14}
}

@inproceedings{10.1145/3544548.3580871,
author = {Namnakani, Omar and Abdrabou, Yasmeen and Grizou, Jonathan and Esteves, Augusto and Khamis, Mohamed},
title = {Comparing Dwell time, Pursuits and Gaze Gestures for Gaze Interaction on Handheld Mobile Devices},
year = {2023},
isbn = {9781450394215},
publisher = {Association for Computing Machinery},
address = {New York, NY, USA},
url = {https://doi.org/10.1145/3544548.3580871},
doi = {10.1145/3544548.3580871},
booktitle = {Proceedings of the 2023 CHI Conference on Human Factors in Computing Systems},
location = {Hamburg, Germany},
series = {CHI '23}
}

@inproceedings{10.1145/3706598.3713092,
author = {Namnakani, Omar and Abdrabou, Yasmeen and Grizou, Jonathan and Khamis, Mohamed},
title = {Stretch Gaze Targets Out: Experimenting with Target Sizes for Gaze-Enabled Interfaces on Mobile Devices},
year = {2025},
isbn = {979-8-4007-1394-1/25/04},
publisher = {Association for Computing Machinery},
address = {New York, NY, USA},
url = {https://doi.org/10.1145/3706598.3713092},
doi = {10.1145/3706598.3713092},
booktitle = {Proceedings of the 2025 CHI Conference on Human Factors in Computing Systems},
location = {Yokohama, Japan},
series = {CHI '25}
}

@article{lei2023dynamicread,
  title={DynamicRead: Exploring Robust Gaze Interaction Methods for Reading on Handheld Mobile Devices under Dynamic Conditions},
  author={Lei, Yaxiong and Wang, Yuheng and Caslin, Tyler and Wisowaty, Alexander and Zhu, Xu and Khamis, Mohamed and Ye, Juan},
  journal={Proceedings of the ACM on Human-Computer Interaction},
  volume={7},
  number={ETRA},
  pages={1--17},
  year={2023},
  publisher={ACM New York, NY, USA}
}

@misc{seeso,
  author = "Eyedid",
  howpublished = {Webpage},
  title={Eyedid SDK, eye-tracking software for your mobile devices},
  url={https://sdk.eyedid.ai},
  lastaccessed=  "September 15, 2025",
  year = {2023}
}

@article{Møllenbach_Hansen_Lillholm_2013, title={Eye Movements in Gaze Interaction}, volume={6}, url={https://bop.unibe.ch/JEMR/article/view/2354}, DOI={10.16910/jemr.6.2.1}, abstractNote={Gaze as a sole input modality must support complex navigation and selection tasks. Gaze interaction combines specific eye movements and graphic display objects (GDOs). This paper suggests a unifying taxonomy of gaze interaction principles. The taxonomy deals with three types of eye movements: fixations, saccades and smooth pursuits and three types of GDOs: static, dynamic, or absent. This taxonomy is qualified through related research and is the first main contribution of this paper. The second part of the paper offers an experimental exploration of single stroke gaze gestures (SSGG). The main findings suggest (1) that different lengths of SSGG can be used for interaction, (2) that GDOs are not necessary for successful completion, and (3) that SSGG are comparable to dwell time selection.}, number={2}, journal={Journal of Eye Movement Research}, author={Møllenbach, Emilie and Hansen, John Paulin and Lillholm, Martin}, year={2013}, month={May} }

@inproceedings{10.1145/2807442.2807499,
author = {Esteves, Augusto and Velloso, Eduardo and Bulling, Andreas and Gellersen, Hans},
title = {Orbits: Gaze Interaction for Smart Watches Using Smooth Pursuit Eye Movements},
year = {2015},
isbn = {9781450337793},
publisher = {Association for Computing Machinery},
address = {New York, NY, USA},
url = {https://doi.org/10.1145/2807442.2807499},
doi = {10.1145/2807442.2807499},
booktitle = {Proceedings of the 28th Annual ACM Symposium on User Interface Software \& Technology},
pages = {457–466},
numpages = {10},
location = {Charlotte, NC, USA},
series = {UIST '15}
}

@inproceedings{10.1145/2168556.2168579,
author = {Heikkil\"{a}, Henna and R\"{a}ih\"{a}, Kari-Jouko},
title = {Simple Gaze Gestures and the Closure of the Eyes as an Interaction Technique},
year = {2012},
isbn = {9781450312219},
publisher = {Association for Computing Machinery},
address = {New York, NY, USA},
url = {https://doi.org/10.1145/2168556.2168579},
doi = {10.1145/2168556.2168579},
booktitle = {Proceedings of the Symposium on Eye Tracking Research and Applications},
pages = {147–154},
numpages = {8},
location = {Santa Barbara, California},
series = {ETRA '12}
}

@inproceedings{10.1145/1378063.1378122,
author = {Drewes, Heiko and De Luca, Alexander and Schmidt, Albrecht},
title = {Eye-Gaze Interaction for Mobile Phones},
year = {2007},
isbn = {9781595938190},
publisher = {Association for Computing Machinery},
address = {New York, NY, USA},
url = {https://doi.org/10.1145/1378063.1378122},
doi = {10.1145/1378063.1378122},
booktitle = {Proceedings of the 4th International Conference on Mobile Technology, Applications, and Systems and the 1st International Symposium on Computer Human Interaction in Mobile Technology},
pages = {364–371},
numpages = {8},
location = {Singapore},
series = {Mobility '07}
}

@inproceedings{10.1145/1743666.1743710,
author = {M\o{}llenbach, Emilie and Lillholm, Martin and Gail, Alastair and Hansen, John Paulin},
title = {Single Gaze Gestures},
year = {2010},
isbn = {9781605589947},
publisher = {Association for Computing Machinery},
address = {New York, NY, USA},
url = {https://doi.org/10.1145/1743666.1743710},
doi = {10.1145/1743666.1743710},
booktitle = {Proceedings of the 2010 Symposium on Eye-Tracking Research \& Applications},
pages = {177–180},
numpages = {4},
location = {Austin, Texas},
series = {ETRA '10}
}

@inbook{10.1145/3419249.3420122,
author = {Fernandez, Misahael and Mathis, Florian and Khamis, Mohamed},
title = {GazeWheels: Comparing Dwell-Time Feedback and Methods for Gaze Input},
year = {2020},
isbn = {9781450375795},
publisher = {Association for Computing Machinery},
address = {New York, NY, USA},
url = {https://doi.org/10.1145/3419249.3420122},
booktitle = {Proceedings of the 11th Nordic Conference on Human-Computer Interaction: Shaping Experiences, Shaping Society},
articleno = {41},
numpages = {6}
}

@inproceedings{10.1145/2493190.2493231,
author = {von Zezschwitz, Emanuel and Dunphy, Paul and De Luca, Alexander},
title = {Patterns in the wild: a field study of the usability of pattern and pin-based authentication on mobile devices},
year = {2013},
isbn = {9781450322737},
publisher = {Association for Computing Machinery},
address = {New York, NY, USA},
url = {https://doi.org/10.1145/2493190.2493231},
doi = {10.1145/2493190.2493231},
booktitle = {Proceedings of the 15th International Conference on Human-Computer Interaction with Mobile Devices and Services},
pages = {261–270},
numpages = {10},
location = {Munich, Germany},
series = {MobileHCI '13}
}

@inproceedings{10.1145/97243.97246,
author = {Jacob, Robert J. K.},
title = {What You Look at is What You Get: Eye Movement-Based Interaction Techniques},
year = {1990},
isbn = {0201509326},
publisher = {Association for Computing Machinery},
address = {New York, NY, USA},
url = {https://doi.org/10.1145/97243.97246},
doi = {10.1145/97243.97246},
booktitle = {Proceedings of the SIGCHI Conference on Human Factors in Computing Systems},
pages = {11–18},
numpages = {8},
location = {Seattle, Washington, USA},
series = {CHI '90}
}

@inproceedings{kong2021eyemu,
  title={EyeMU Interactions: Gaze+ IMU Gestures on Mobile Devices},
  author={Kong, Andy and Ahuja, Karan and Goel, Mayank and Harrison, Chris},
  booktitle={Proceedings of the 2021 International Conference on Multimodal Interaction},
  pages={577--585},
  year={2021}
}

@inproceedings{10.1145/2642918.2647397,
author = {Pfeuffer, Ken and Alexander, Jason and Chong, Ming Ki and Gellersen, Hans},
title = {Gaze-Touch: Combining Gaze with Multi-Touch for Interaction on the Same Surface},
year = {2014},
isbn = {9781450330695},
publisher = {Association for Computing Machinery},
address = {New York, NY, USA},
url = {https://doi.org/10.1145/2642918.2647397},
doi = {10.1145/2642918.2647397},
booktitle = {Proceedings of the 27th Annual ACM Symposium on User Interface Software and Technology},
pages = {509–518},
numpages = {10},
location = {Honolulu, Hawaii, USA},
series = {UIST '14}
}

@inproceedings{10.1145/2984511.2984514,
author = {Pfeuffer, Ken and Gellersen, Hans},
title = {Gaze and Touch Interaction on Tablets},
year = {2016},
isbn = {9781450341899},
publisher = {Association for Computing Machinery},
address = {New York, NY, USA},
url = {https://doi.org/10.1145/2984511.2984514},
doi = {10.1145/2984511.2984514},
booktitle = {Proceedings of the 29th Annual Symposium on User Interface Software and Technology},
pages = {301–311},
numpages = {11},
location = {Tokyo, Japan},
series = {UIST '16}
}

@article{10.1145/123078.128728,
author = {Jacob, Robert J. K.},
title = {The Use of Eye Movements in Human-Computer Interaction Techniques: What You Look at is What You Get},
year = {1991},
issue_date = {April 1991},
publisher = {Association for Computing Machinery},
address = {New York, NY, USA},
volume = {9},
number = {2},
issn = {1046-8188},
url = {https://doi.org/10.1145/123078.128728},
doi = {10.1145/123078.128728},
journal = {ACM Trans. Inf. Syst.},
month = {apr},
pages = {152–169},
numpages = {18}
}

@inproceedings{10.1145/3206505.3206522,
author = {Khamis, Mohamed and Oechsner, Carl and Alt, Florian and Bulling, Andreas},
title = {VRpursuits: Interaction in Virtual Reality Using Smooth Pursuit Eye Movements},
year = {2018},
isbn = {9781450356169},
publisher = {Association for Computing Machinery},
address = {New York, NY, USA},
url = {https://doi.org/10.1145/3206505.3206522},
doi = {10.1145/3206505.3206522},
booktitle = {Proceedings of the 2018 International Conference on Advanced Visual Interfaces},
articleno = {18},
numpages = {8},
location = {Castiglione della Pescaia, Grosseto, Italy},
series = {AVI '18}
}

@inproceedings{VidalPusuits2013,
author = {Vidal, M\'{e}lodie and Pfeuffer, Ken and Bulling, Andreas and Gellersen, Hans W.},
title = {Pursuits: Eye-Based Interaction with Moving Targets},
year = {2013},
isbn = {9781450319522},
publisher = {Association for Computing Machinery},
address = {New York, NY, USA},
url = {https://doi.org/10.1145/2468356.2479632},
doi = {10.1145/2468356.2479632},
booktitle = {CHI '13 Extended Abstracts on Human Factors in Computing Systems},
pages = {3147–3150},
numpages = {4},
location = {Paris, France},
series = {CHI EA '13}
}

@inproceedings{AbdrabouJustGaze2019,
author = {Abdrabou, Yasmeen and Khamis, Mohamed and Eisa, Rana Mohamed and Ismail, Sherif and Elmougy, Amrl},
title = {Just Gaze and Wave: Exploring the Use of Gaze and Gestures for Shoulder-Surfing Resilient Authentication},
year = {2019},
isbn = {9781450367097},
publisher = {Association for Computing Machinery},
address = {New York, NY, USA},
url = {https://doi.org/10.1145/3314111.3319837},
doi = {10.1145/3314111.3319837},
booktitle = {Proceedings of the 11th ACM Symposium on Eye Tracking Research \& Applications},
articleno = {29},
numpages = {10},
location = {Denver, Colorado},
series = {ETRA '19}
}

@inproceedings{Gaze+pinch,
author = {Pfeuffer, Ken and Mayer, Benedikt and Mardanbegi, Diako and Gellersen, Hans},
title = {Gaze + Pinch Interaction in Virtual Reality},
year = {2017},
isbn = {9781450354868},
publisher = {Association for Computing Machinery},
address = {New York, NY, USA},
url = {https://doi.org/10.1145/3131277.3132180},
doi = {10.1145/3131277.3132180},
booktitle = {Proceedings of the 5th Symposium on Spatial User Interaction},
pages = {99–108},
numpages = {10},
location = {Brighton, United Kingdom},
series = {SUI '17}
}

@article{esteves2020comparing,
title = {Comparing selection mechanisms for gaze input techniques in head-mounted displays},
journal = {International Journal of Human-Computer Studies},
volume = {139},
pages = {102414},
year = {2020},
issn = {1071-5819},
doi = {https://doi.org/10.1016/j.ijhcs.2020.102414},
url = {https://www.sciencedirect.com/science/article/pii/S1071581920300185},
author = {Augusto Esteves and Yonghwan Shin and Ian Oakley}
}

@inproceedings{10.1145/2556288.2557312,
author = {Ng, Alexander and Brewster, Stephen A. and Williamson, John H.},
title = {Investigating the effects of encumbrance on one- and two- handed interactions with mobile devices},
year = {2014},
isbn = {9781450324731},
publisher = {Association for Computing Machinery},
address = {New York, NY, USA},
url = {https://doi.org/10.1145/2556288.2557312},
doi = {10.1145/2556288.2557312},
booktitle = {Proceedings of the SIGCHI Conference on Human Factors in Computing Systems},
pages = {1981–1990},
numpages = {10},
location = {Toronto, Ontario, Canada},
series = {CHI '14}
}

@inproceedings{10.1145/2628363.2628382,
author = {Ng, Alexander and Williamson, John H. and Brewster, Stephen A.},
title = {Comparing evaluation methods for encumbrance and walking on interaction with touchscreen mobile devices},
year = {2014},
isbn = {9781450330046},
publisher = {Association for Computing Machinery},
address = {New York, NY, USA},
url = {https://doi.org/10.1145/2628363.2628382},
doi = {10.1145/2628363.2628382},
booktitle = {Proceedings of the 16th International Conference on Human-Computer Interaction with Mobile Devices \& Services},
pages = {23–32},
numpages = {10},
location = {Toronto, ON, Canada},
series = {MobileHCI '14}
}

@inproceedings{10.1145/2414536.2414609,
author = {Penkar, Abdul Moiz and Lutteroth, Christof and Weber, Gerald},
title = {Designing for the Eye: Design Parameters for Dwell in Gaze Interaction},
year = {2012},
isbn = {9781450314381},
publisher = {Association for Computing Machinery},
address = {New York, NY, USA},
url = {https://doi.org/10.1145/2414536.2414609},
doi = {10.1145/2414536.2414609},
booktitle = {Proceedings of the 24th Australian Computer-Human Interaction Conference},
pages = {479–488},
numpages = {10},
location = {Melbourne, Australia},
series = {OzCHI '12}
}

@article{doi:10.1177/154193120605000909,
author = {Sandra G. Hart},
title ={Nasa-Task Load Index (NASA-TLX); 20 Years Later},
journal = {Proceedings of the Human Factors and Ergonomics Society Annual Meeting},
volume = {50},
number = {9},
pages = {904-908},
year = {2006},
doi = {10.1177/154193120605000909},

URL = { 
        https://doi.org/10.1177/154193120605000909
    
},
eprint = { 
        https://doi.org/10.1177/154193120605000909
    
}
}

@inproceedings{10.1145/3152771.3156161,
author = {Sarsenbayeva, Zhanna and van Berkel, Niels and Luo, Chu and Kostakos, Vassilis and Goncalves, Jorge},
title = {Challenges of situational impairments during interaction with mobile devices},
year = {2017},
isbn = {9781450353793},
publisher = {Association for Computing Machinery},
address = {New York, NY, USA},
url = {https://doi.org/10.1145/3152771.3156161},
doi = {10.1145/3152771.3156161},
booktitle = {Proceedings of the 29th Australian Conference on Computer-Human Interaction},
pages = {477–481},
numpages = {5},
location = {Brisbane, Queensland, Australia},
series = {OzCHI '17}
}

@article{wobbrock2019situationally,
  title={Situationally-induced impairments and disabilities},
  author={Wobbrock, Jacob O},
  journal={Web Accessibility: A Foundation for Research},
  pages={59--92},
  year={2019},
  publisher={Springer}
}

@inproceedings{wobbrock2006future,
  title={The future of mobile device research in HCI},
  author={Wobbrock, Jacob O},
  booktitle={CHI 2006 workshop proceedings: what is the next generation of human-computer interaction},
  pages={131--134},
  year={2006}
}

@inproceedings{10.1145/1978942.1978963,
author = {Wobbrock, Jacob O. and Findlater, Leah and Gergle, Darren and Higgins, James J.},
title = {The aligned rank transform for nonparametric factorial analyses using only anova procedures},
year = {2011},
isbn = {9781450302289},
publisher = {Association for Computing Machinery},
address = {New York, NY, USA},
url = {https://doi.org/10.1145/1978942.1978963},
doi = {10.1145/1978942.1978963},
booktitle = {Proceedings of the SIGCHI Conference on Human Factors in Computing Systems},
pages = {143–146},
numpages = {4},
location = {Vancouver, BC, Canada},
series = {CHI '11}
}

@inproceedings{10.1145/1409240.1409253,
author = {Kane, Shaun K. and Wobbrock, Jacob O. and Smith, Ian E.},
title = {Getting off the treadmill: evaluating walking user interfaces for mobile devices in public spaces},
year = {2008},
isbn = {9781595939524},
publisher = {Association for Computing Machinery},
address = {New York, NY, USA},
url = {https://doi.org/10.1145/1409240.1409253},
doi = {10.1145/1409240.1409253},
booktitle = {Proceedings of the 10th International Conference on Human Computer Interaction with Mobile Devices and Services},
pages = {109–118},
numpages = {10},
location = {Amsterdam, The Netherlands},
series = {MobileHCI '08}
}

@article{saulynas2022putting,
  title={Putting situational impairments in context: developing guidance for situational impairments and severely constraining situational impairments by examining parallel domains},
  author={Saulynas, Sidas and Burgee, Lawrence and Bendigeri, Apoorva and Kuber, Ravi},
  journal={Universal Access in the Information Society},
  pages={1--26},
  year={2022},
  publisher={Springer}
}

@inproceedings{10.1145/1983302.1983303,
author = {Stellmach, Sophie and Stober, Sebastian and N\"{u}rnberger, Andreas and Dachselt, Raimund},
title = {Designing gaze-supported multimodal interactions for the exploration of large image collections},
year = {2011},
isbn = {9781450306805},
publisher = {Association for Computing Machinery},
address = {New York, NY, USA},
url = {https://doi.org/10.1145/1983302.1983303},
doi = {10.1145/1983302.1983303},
booktitle = {Proceedings of the 1st Conference on Novel Gaze-Controlled Applications},
articleno = {1},
numpages = {8},
location = {Karlskrona, Sweden},
series = {NGCA '11}
}

@inproceedings{10.1145/1054972.1054994,
author = {Fono, David and Vertegaal, Roel},
title = {EyeWindows: evaluation of eye-controlled zooming windows for focus selection},
year = {2005},
isbn = {1581139985},
publisher = {Association for Computing Machinery},
address = {New York, NY, USA},
url = {https://doi.org/10.1145/1054972.1054994},
doi = {10.1145/1054972.1054994},
booktitle = {Proceedings of the SIGCHI Conference on Human Factors in Computing Systems},
pages = {151–160},
numpages = {10},
location = {Portland, Oregon, USA},
series = {CHI '05}
}

@inproceedings{10.1145/2207676.2208709,
author = {Stellmach, Sophie and Dachselt, Raimund},
title = {Look \& touch: gaze-supported target acquisition},
year = {2012},
isbn = {9781450310154},
publisher = {Association for Computing Machinery},
address = {New York, NY, USA},
url = {https://doi.org/10.1145/2207676.2208709},
doi = {10.1145/2207676.2208709},
booktitle = {Proceedings of the SIGCHI Conference on Human Factors in Computing Systems},
pages = {2981–2990},
numpages = {10},
location = {Austin, Texas, USA},
series = {CHI '12}
}

@article{goncalves2017tapping,
  title={Tapping task performance on smartphones in cold temperature},
  author={Goncalves, Jorge and Sarsenbayeva, Zhanna and van Berkel, Niels and Luo, Chu and Hosio, Simo and Risanen, Sirkka and Rintam{\"a}ki, Hannu and Kostakos, Vassilis},
  journal={Interacting with Computers},
  volume={29},
  number={3},
  pages={355--367},
  year={2017},
  publisher={Oxford University Press}
}

@inproceedings{salvucci2000intelligent,
  title={Intelligent gaze-added interfaces},
  author={Salvucci, Dario D and Anderson, John R},
  booktitle={Proceedings of the SIGCHI conference on Human factors in computing systems},
  pages={273--280},
  year={2000}
}

@inproceedings{10.1145/3490099.3511103,
author = {Zhao, Maozheng and Huang, Henry and Li, Zhi and Liu, Rui and Cui, Wenzhe and Toshniwal, Kajal and Goel, Ananya and Wang, Andrew and Zhao, Xia and Rashidian, Sina and Baig, Furqan and Phi, Khiem and Zhai, Shumin and Ramakrishnan, IV and Wang, Fusheng and Bi, Xiaojun},
title = {EyeSayCorrect: Eye Gaze and Voice Based Hands-free Text Correction for Mobile Devices},
year = {2022},
isbn = {9781450391443},
publisher = {Association for Computing Machinery},
address = {New York, NY, USA},
url = {https://doi.org/10.1145/3490099.3511103},
doi = {10.1145/3490099.3511103},
booktitle = {Proceedings of the 27th International Conference on Intelligent User Interfaces},
pages = {470–482},
numpages = {13},
location = {Helsinki, Finland},
series = {IUI '22}
}

@inproceedings{10.1145/332040.332445,
author = {Sibert, Linda E. and Jacob, Robert J. K.},
title = {Evaluation of eye gaze interaction},
year = {2000},
isbn = {1581132166},
publisher = {Association for Computing Machinery},
address = {New York, NY, USA},
url = {https://doi.org/10.1145/332040.332445},
doi = {10.1145/332040.332445},
booktitle = {Proceedings of the SIGCHI Conference on Human Factors in Computing Systems},
pages = {281–288},
numpages = {8},
location = {The Hague, The Netherlands},
series = {CHI '00}
}

@article{ng2011effects,
  title={The effects of encumbrance on mobile gesture interactions},
  author={Ng, Alexander and Brewster, Stephen and Crossan, Andrew},
  year={2011}
}

@InProceedings{Ng2013Encumbrance,
  author    = {Ng, Alexander and Brewster, Stephen A. and Williamson, John},
  editor    = {Kotz{\'e}, Paula and Marsden, Gary and Lindgaard, Gitte and Wesson, Janet and Winckler, Marco},
  title     = {The Impact of Encumbrance on Mobile Interactions},
  booktitle = {Human-Computer Interaction -- INTERACT 2013},
  year      = {2013},
  publisher = {Springer Berlin Heidelberg},
  address   = {Berlin, Heidelberg},
  pages     = {92--109},
  isbn      = {978-3-642-40477-1},
  doi       = {10.1007/978-3-642-40477-1_6}
}

@inproceedings{10.1145/3123021.3123033,
author = {Dobbelstein, David and Haas, Gabriel and Rukzio, Enrico},
title = {The effects of mobility, encumbrance, and (non-)dominant hand on interaction with smartwatches},
year = {2017},
isbn = {9781450351881},
publisher = {Association for Computing Machinery},
address = {New York, NY, USA},
url = {https://doi.org/10.1145/3123021.3123033},
doi = {10.1145/3123021.3123033},
booktitle = {Proceedings of the 2017 ACM International Symposium on Wearable Computers},
pages = {90–93},
numpages = {4},
location = {Maui, Hawaii},
series = {ISWC '17}
}

@inproceedings{10.1145/2785830.2785853,
author = {Ng, Alexander and Williamson, John and Brewster, Stephen},
title = {The Effects of Encumbrance and Mobility on Touch-Based Gesture Interactions for Mobile Phones},
year = {2015},
isbn = {9781450336529},
publisher = {Association for Computing Machinery},
address = {New York, NY, USA},
url = {https://doi.org/10.1145/2785830.2785853},
doi = {10.1145/2785830.2785853},
booktitle = {Proceedings of the 17th International Conference on Human-Computer Interaction with Mobile Devices and Services},
pages = {536–546},
numpages = {11},
location = {Copenhagen, Denmark},
series = {MobileHCI '15}
}

@inproceedings{10.1145/3242969.3243028,
author = {Tung, Ying-Chao and Goel, Mayank and Zinda, Isaac and Wobbrock, Jacob O.},
title = {RainCheck: Overcoming Capacitive Interference Caused by Rainwater on Smartphones},
year = {2018},
isbn = {9781450356923},
publisher = {Association for Computing Machinery},
address = {New York, NY, USA},
url = {https://doi.org/10.1145/3242969.3243028},
doi = {10.1145/3242969.3243028},
booktitle = {Proceedings of the 20th ACM International Conference on Multimodal Interaction},
pages = {464–471},
numpages = {8},
location = {Boulder, CO, USA},
series = {ICMI '18}
}

@inproceedings{10.1145/3204493.3208344,
author = {Rajanna, Vijay and Hammond, Tracy},
title = {A gaze gesture-based paradigm for situational impairments, accessibility, and rich interactions},
year = {2018},
isbn = {9781450357067},
publisher = {Association for Computing Machinery},
address = {New York, NY, USA},
url = {https://doi.org/10.1145/3204493.3208344},
doi = {10.1145/3204493.3208344},
booktitle = {Proceedings of the 2018 ACM Symposium on Eye Tracking Research \& Applications},
articleno = {102},
numpages = {3},
location = {Warsaw, Poland},
series = {ETRA '18}
}

@inproceedings{10.1145/3319499.3330292,
author = {Wobbrock, Jacob O.},
title = {Situationally aware mobile devices for overcoming situational impairments},
year = {2019},
isbn = {9781450367455},
publisher = {Association for Computing Machinery},
address = {New York, NY, USA},
url = {https://doi.org/10.1145/3319499.3330292},
doi = {10.1145/3319499.3330292},
booktitle = {Proceedings of the ACM SIGCHI Symposium on Engineering Interactive Computing Systems},
articleno = {1},
numpages = {18},
location = {Valencia, Spain},
series = {EICS '19}
}

@inproceedings{10.1145/2971648.2971734,
author = {Sarsenbayeva, Zhanna and Goncalves, Jorge and Garc\'{\i}a, Juan and Klakegg, Simon and Rissanen, Sirkka and Rintam\"{a}ki, Hannu and Hannu, Jari and Kostakos, Vassilis},
title = {Situational impairments to mobile interaction in cold environments},
year = {2016},
isbn = {9781450344616},
publisher = {Association for Computing Machinery},
address = {New York, NY, USA},
url = {https://doi.org/10.1145/2971648.2971734},
doi = {10.1145/2971648.2971734},
booktitle = {Proceedings of the 2016 ACM International Joint Conference on Pervasive and Ubiquitous Computing},
pages = {85–96},
numpages = {12},
location = {Heidelberg, Germany},
series = {UbiComp '16}
}

@article{10.1145/3577013,
author = {Akpinar, Elgin and Ye\c{s}ilada, Yeliz and Karag\"{o}z, Pinar},
title = {Effect of Context on Smartphone Users’ Typing Performance in the Wild},
year = {2023},
issue_date = {June 2023},
publisher = {Association for Computing Machinery},
address = {New York, NY, USA},
volume = {30},
number = {3},
issn = {1073-0516},
url = {https://doi.org/10.1145/3577013},
doi = {10.1145/3577013},
journal = {ACM Trans. Comput.-Hum. Interact.},
month = jun,
articleno = {36},
numpages = {44}
}

@inproceedings{10.1145/3173574.3174208,
author = {Singh, Gaganpreet and Delamare, William and Irani, Pourang},
title = {D-SWIME: A Design Space for Smartwatch Interaction Techniques Supporting Mobility and Encumbrance},
year = {2018},
isbn = {9781450356206},
publisher = {Association for Computing Machinery},
address = {New York, NY, USA},
url = {https://doi.org/10.1145/3173574.3174208},
doi = {10.1145/3173574.3174208},
booktitle = {Proceedings of the 2018 CHI Conference on Human Factors in Computing Systems},
pages = {1–13},
numpages = {13},
location = {Montreal QC, Canada},
series = {CHI '18}
}

@article{10.1145/3386370,
author = {Akpinar, Elgin and Ye\c{s}ilada, Yeliz and Temizer, Selim},
title = {The Effect of Context on Small Screen and Wearable Device Users’ Performance - A Systematic Review},
year = {2020},
issue_date = {May 2021},
publisher = {Association for Computing Machinery},
address = {New York, NY, USA},
volume = {53},
number = {3},
issn = {0360-0300},
url = {https://doi.org/10.1145/3386370},
doi = {10.1145/3386370},
journal = {ACM Comput. Surv.},
month = may,
articleno = {52},
numpages = {44}
}

@misc{apple_tracking,
    author    = {{Apple Inc.}},
    title     = {Control iPhone with the movement of your eyes},
    url       = {https://support.apple.com/en-gb/guide/iphone/iph66057d0f6/ios},
    year= {2024},
    note      = {Accessed: 22 September 2025}
}

@Article{Orphanides2017Touchscreen,
  author   = {Andreas K. Orphanides and Chang S. Nam},
  title    = {Touchscreen interfaces in context: A systematic review of research into touchscreens across settings, populations, and implementations},
  journal  = {Applied Ergonomics},
  year     = {2017},
  volume   = {61},
  pages    = {116--143},
  issn     = {0003-6870},
  doi      = {10.1016/j.apergo.2017.01.013},
  url      = {https://www.sciencedirect.com/science/article/pii/S0003687017300212}
}

@article{10.1145/3591133,
author = {Alsakar, Noora and Abdrabou, Yasmeen and Stumpf, Simone and Khamis, Mohamed},
title = {Investigating Privacy Perceptions and Subjective Acceptance of Eye Tracking on Handheld Mobile Devices},
year = {2023},
issue_date = {May 2023},
publisher = {Association for Computing Machinery},
address = {New York, NY, USA},
volume = {7},
number = {ETRA},
url = {https://doi.org/10.1145/3591133},
doi = {10.1145/3591133},
journal = {Proc. ACM Hum.-Comput. Interact.},
month = may,
articleno = {164},
numpages = {16}
}

@InProceedings{Pfeuffer2015GazeTouch,
  author    = {Pfeuffer, Ken and Alexander, Jason and Gellersen, Hans},
  editor    = {Abascal, Julio and Barbosa, Simone and Fetter, Mirko and Gross, Tom and Palanque, Philippe and Winckler, Marco},
  title     = {Gaze+touch vs. Touch: What's the Trade-off When Using Gaze to Extend Touch to Remote Displays?},
  booktitle = {Human-Computer Interaction -- INTERACT 2015},
  year      = {2015},
  publisher = {Springer International Publishing},
  address   = {Cham},
  pages     = {349--367},
  isbn      = {978-3-319-22668-2},
  doi       = {10.1007/978-3-319-22668-2_27}
}

@inproceedings{10.1145/2858036.2858201,
author = {Pfeuffer, Ken and Alexander, Jason and Gellersen, Hans},
title = {Partially-indirect Bimanual Input with Gaze, Pen, and Touch for Pan, Zoom, and Ink Interaction},
year = {2016},
isbn = {9781450333627},
publisher = {Association for Computing Machinery},
address = {New York, NY, USA},
url = {https://doi.org/10.1145/2858036.2858201},
doi = {10.1145/2858036.2858201},
booktitle = {Proceedings of the 2016 CHI Conference on Human Factors in Computing Systems},
pages = {2845–2856},
numpages = {12},
location = {San Jose, California, USA},
series = {CHI '16}
}

@InBook{Hansen2003DwellTyping,
  author    = {Hansen, John Paulin and Johansen, Anders Sewerin and Hansen, Dan Witzner and Itoh, Kenji and Mashino, Satoru},
  title     = {Command Without a Click: Dwell Time Typing by Mouse and Gaze Selections},
  booktitle = {Human-Computer Interaction - INTERACT'03},
  editor    = {Rauterberg, M.},
  year      = {2003},
  pages     = {121--128},
  publisher = {IOS Press},
  language  = {English},
  note      = {10th International Conference on Human-Computer Interaction, HCI International 2003; Conference date: 22--27 June 2003},
  url       = {http://www.hci-international.org/index.php?module=conference&CF_op=view&CF_id=3}
}

@software{qcamap,
  author = {{University of Klagenfurt}},
  title = {{QCAmap}: Qualitative Content Analysis Program},
  year = {2013},
  url = {https://www.qcamap.org},
  note = {accessed June 20, 2025},
}

@inproceedings{10.1145/3706598.3713492,
author = {Li, Tinghui and Velloso, Eduardo and Withana, Anusha and Sarsenbayeva, Zhanna},
title = {Estimating the Effects of Encumbrance and Walking on Mixed Reality Interaction},
year = {2025},
isbn = {9798400713941},
publisher = {Association for Computing Machinery},
address = {New York, NY, USA},
url = {https://doi.org/10.1145/3706598.3713492},
doi = {10.1145/3706598.3713492},
booktitle = {Proceedings of the 2025 CHI Conference on Human Factors in Computing Systems},
articleno = {1153},
numpages = {24},
location = {
},
series = {CHI '25}
}

@inproceedings{10.1145/3613904.3642712,
author = {Brailsford, Joe and Vetere, Frank and Velloso, Eduardo},
title = {Exploring the Association between Moral Foundations and Judgements of AI Behaviour},
year = {2024},
isbn = {9798400703300},
publisher = {Association for Computing Machinery},
address = {New York, NY, USA},
url = {https://doi.org/10.1145/3613904.3642712},
doi = {10.1145/3613904.3642712},
booktitle = {Proceedings of the 2024 CHI Conference on Human Factors in Computing Systems},
articleno = {284},
numpages = {15},
location = {Honolulu, HI, USA},
series = {CHI '24}
}

@article{JSSv080i01,
 title={brms: An R Package for Bayesian Multilevel Models Using Stan},
 volume={80},
 url={https://www.jstatsoft.org/index.php/jss/article/view/v080i01},
 doi={10.18637/jss.v080.i01},
 number={1},
 journal={Journal of Statistical Software},
 author={Bürkner, Paul-Christian},
 year={2017},
 pages={1–28}
}

@inproceedings{10.1145/3613904.3642919,
author = {Wadinambiarachchi, Samangi and Kelly, Ryan M. and Pareek, Saumya and Zhou, Qiushi and Velloso, Eduardo},
title = {The Effects of Generative AI on Design Fixation and Divergent Thinking},
year = {2024},
isbn = {9798400703300},
publisher = {Association for Computing Machinery},
address = {New York, NY, USA},
url = {https://doi.org/10.1145/3613904.3642919},
doi = {10.1145/3613904.3642919},
booktitle = {Proceedings of the 2024 CHI Conference on Human Factors in Computing Systems},
articleno = {380},
numpages = {18},
location = {Honolulu, HI, USA},
series = {CHI '24}
}

@book{russo2003statistics,
  author    = {Russo, Riccardo},
  title     = {Statistics for the Behavioural Sciences: An Introduction},
  year      = {2003},
  edition   = {1st},
  publisher = {Psychology Press},
  doi       = {10.4324/9780203641576},
  address   = {Hove, East Sussex, UK},
}

@article{wagenmakers2011why,
  author    = {Wagenmakers, Eric-Jan and Wetzels, Ruud and Borsboom, Denny and van der Maas, Han L. J.},
  title     = {Why Psychologists Must Change the Way They Analyze Their Data: The Case of Psi: Comment on Bem (2011)},
  journal   = {Journal of Personality and Social Psychology},
  year      = {2011},
  volume    = {100},
  number    = {3},
  pages     = {426--432},
  doi       = {10.1037/a0022790},
}

@article{Tornblom2025,
  author    = {Törnblom, Madelene and Rönkkö, Kari and Ådahl, Kerstin and Karlsson, Staffan and Olsson Möller, Ulrika and Nivestam, Anna},
  title     = {Older persons’ experiences with wearable sensor-based fall risk screening in free-living conditions -- a qualitative study},
  year      = {2025},
  journal   = {BMC Geriatrics},
  volume    = {25},
  number    = {1},
  articleno = {426},
  doi       = {10.1186/s12877-025-06100-7},
  issn      = {1471-2318},
  publisher = {Springer Nature},
  address   = {London, United Kingdom},
  url       = {https://doi.org/10.1186/s12877-025-06100-7}
}

@inproceedings{10.1145/3279778.3279791,
author = {Lehmann, Florian and Kipp, Michael},
title = {How to Hold Your Phone When Tapping: A Comparative Study of Performance, Precision, and Errors},
year = {2018},
isbn = {9781450356947},
publisher = {Association for Computing Machinery},
address = {New York, NY, USA},
url = {https://doi.org/10.1145/3279778.3279791},
doi = {10.1145/3279778.3279791},
booktitle = {Proceedings of the 2018 ACM International Conference on Interactive Surfaces and Spaces},
pages = {115–127},
numpages = {13},
location = {Tokyo, Japan},
series = {ISS '18}
}

@inproceedings{Karlson2006,
  author       = {Karlson, Amy K. and Bederson, Benjamin B. and Contreras-Vidal, Jose L.},
  title        = {Studies in One-Handed Mobile Design: Habit, Desire and Agility},
  booktitle    = {Proceedings of the 4th ERCIM Workshop on User Interfaces for All (UI4ALL)},
  year         = {2006},
  pages        = {1--10},
  publisher    = {Citeseer},
  address      = {Princeton, NJ, USA}
}

@article{Bhattacharya2019,
  author    = {Bhattacharya, Sudip and Bashar, Md Abu and Srivastava, Abhay and Singh, Amarjeet},
  title     = {NOMOPHOBIA: NO MObile PHone PhoBIA},
  journal   = {Journal of Family Medicine and Primary Care},
  year      = {2019},
  volume    = {8},
  number    = {4},
  pages     = {1297--1300},
  doi       = {10.4103/jfmpc.jfmpc_71_19},
  issn      = {2249-4863},
  publisher = {Wolters Kluwer Medknow Publications},
  address   = {Mumbai, India},
  url       = {https://doi.org/10.4103/jfmpc.jfmpc_71_19}
}

@inproceedings{10.1145/3173574.3173605,
author = {Le, Huy Viet and Mayer, Sven and Bader, Patrick and Henze, Niels},
title = {Fingers' Range and Comfortable Area for One-Handed Smartphone Interaction Beyond the Touchscreen},
year = {2018},
isbn = {9781450356206},
publisher = {Association for Computing Machinery},
address = {New York, NY, USA},
url = {https://doi.org/10.1145/3173574.3173605},
doi = {10.1145/3173574.3173605},
booktitle = {Proceedings of the 2018 CHI Conference on Human Factors in Computing Systems},
pages = {1–12},
numpages = {12},
location = {Montreal QC, Canada},
series = {CHI '18}
}

@article{Vaportzis2017,
  author    = {Vaportzis, Eleftheria and Giatsi Clausen, Maria and Gow, Alan J.},
  title     = {Older Adults' Perceptions of Technology and Barriers to Interacting with Tablet Computers: A Focus Group Study},
  year      = {2017},
  journal   = {Frontiers in Psychology},
  volume    = {8},
  doi       = {10.3389/fpsyg.2017.01687},
  issn      = {1664-1078},
  url       = {https://www.frontiersin.org/articles/10.3389/fpsyg.2017.01687},
  publisher = {Frontiers Media SA},
  address   = {Lausanne, Switzerland},
  articleno = {1687}
}

@inproceedings{10.1145/3338286.3340133,
author = {Kuosmanen, Elina and Kan, Valerii and Vega, Julio and Visuri, Aku and Nishiyama, Yuuki and Dey, Anind K. and Harper, Simon and Ferreira, Denzil},
title = {Challenges of Parkinson's Disease: User Experiences with STOP},
year = {2019},
isbn = {9781450368254},
publisher = {Association for Computing Machinery},
address = {New York, NY, USA},
url = {https://doi.org/10.1145/3338286.3340133},
doi = {10.1145/3338286.3340133},
booktitle = {Proceedings of the 21st International Conference on Human-Computer Interaction with Mobile Devices and Services},
articleno = {22},
numpages = {11},
location = {Taipei, Taiwan},
series = {MobileHCI '19}
}

@article{Wilson2023,
  author    = {Wilson, Gemma and Gates, Jessica R. and Vijaykumar, Santosh and Morgan, Deborah J.},
  title     = {Understanding older adults’ use of social technology and the factors influencing use},
  year      = {2023},
  journal   = {Ageing and Society},
  volume    = {43},
  number    = {1},
  pages     = {222--245},
  doi       = {10.1017/S0144686X21000490},
  publisher = {Cambridge University Press},
  address   = {Cambridge, United Kingdom},
  issn      = {0144-686X},
  url       = {https://doi.org/10.1017/S0144686X21000490}
}

@inproceedings{Taylor2022,
  author       = {Taylor, Tiffany E.},
  title        = {The User’s Experience: Exploring the Impact Our Interactions with Technology Have on Us},
  booktitle    = {Proceedings of the 35th International BCS Human-Computer Interaction Conference},
  pages        = {1--9},
  year         = {2022},
  doi          = {10.14236/ewic/HCI2022.36},
  url          = {https://doi.org/10.14236/ewic/HCI2022.36},
  publisher    = {BCS Learning and Development},
  address      = {Swindon, United Kingdom}
}

@article{Ferreri2021,
  author    = {Ferreri, Nicholas and Mayhorn, Christopher B.},
  title     = {That’s Not What We Expected: Examining Technology Expectations and Malfunctions on Frustration},
  journal   = {Ergonomics in Design: The Quarterly of Human Factors Applications},
  year      = {2021},
  volume    = {31},
  number    = {4},
  pages     = {40--45},
  doi       = {10.1177/10648046211007709},
  url       = {https://doi.org/10.1177/10648046211007709},
  publisher = {SAGE Publications},
  address   = {Thousand Oaks, CA, USA},
  issn      = {1064-8046},
  note      = {Original work published 2023}
}

@inproceedings{10.1145/968363.968390,
author = {Majaranta, P\"{a}ivi and Aula, Anne and R\"{a}ih\"{a}, Kari-Jouko},
title = {Effects of feedback on eye typing with a short dwell time},
year = {2004},
isbn = {1581138253},
publisher = {Association for Computing Machinery},
address = {New York, NY, USA},
url = {https://doi.org/10.1145/968363.968390},
doi = {10.1145/968363.968390},
booktitle = {Proceedings of the 2004 Symposium on Eye Tracking Research \& Applications},
pages = {139–146},
numpages = {8},
location = {San Antonio, Texas},
series = {ETRA '04}
}

@article{Majaranta2006,
  author    = {Majaranta, Päivi and MacKenzie, I. Scott and Aula, Anne and Räihä, Kari-Jouko},
  title     = {Effects of Feedback and Dwell Time on Eye Typing Speed and Accuracy},
  journal   = {Universal Access in the Information Society},
  year      = {2006},
  volume    = {5},
  number    = {2},
  pages     = {199--208},
  doi       = {10.1007/s10209-006-0034-z},
  url       = {https://doi.org/10.1007/s10209-006-0034-z},
  issn      = {1615-5297},
  publisher = {Springer},
  address   = {Berlin, Germany}
}

@incollection{LaViola2015Multimodal,
  author    = {Joseph J. LaViola Jr. and Sarah Buchanan and Corey Pittman},
  title     = {Multimodal Input for Perceptual User Interfaces},
  booktitle = {Interactive Displays: Natural Human-Interface Technologies},
  editor    = {Achintya K. Bhowmik},
  publisher = {John Wiley \& Sons, Ltd},
  year      = {2015},
  pages     = {285--312},
  doi       = {10.1002/9781118706237.ch9}
}

@article{Fernandes2025,
  author    = {Fernandes, Ajoy S. and Schütz, Immo and Murdison, T. Scott and Proulx, Michael J.},
  title     = {Gaze Inputs for Targeting: The Eyes Have It, Not With a Cursor},
  journal   = {International Journal of Human--Computer Interaction},
  year      = {2025},
  volume    = {41},
  number    = {19},
  pages     = {12251--12269},
  doi       = {10.1080/10447318.2025.2453966},
  url       = {https://doi.org/10.1080/10447318.2025.2453966},
  publisher = {Taylor \& Francis},
  address   = {London, United Kingdom}
}

@INPROCEEDINGS{6252501,
  author={Santhiranayagam, Braveena K. and Lai, Daniel T. H. and Jiang, Cancan and Shilton, Alistair and Begg, Rezaul},
  booktitle={The 2012 International Joint Conference on Neural Networks (IJCNN)}, 
  title={Automatic detection of different walking conditions using inertial sensor data}, 
  year={2012},
  volume={},
  number={},
  pages={1-6},
  doi={10.1109/IJCNN.2012.6252501}}

\section{Appendices}
\appendix

% Please add the following required packages to your document preamble:
% \usepackage{multirow}
\begin{table*}[]
\centering
\caption{Likert responses on usability aspects. We explored the impact of enumbrance on each input modality participants' perception. We report the Median, Q1 and Q3 along the results of the Friedman tests to check for singificance.}
\label{tab_likert_significance_all}
\begin{tabular}{llllll}
                                                                                                                   &                                              & \multicolumn{2}{c}{\textbf{Encumbrance Level}}                                          & \multicolumn{2}{c}{\textbf{Friedman}}                                        \\ 
\multicolumn{1}{l}{\textbf{Usability Aspect}}                                                                    & \multicolumn{1}{l}{\textbf{Input Modality}} & \multicolumn{1}{l}{\textbf{Encumbered}}   & \multicolumn{1}{l}{\textbf{Unencumbered}} & \multicolumn{1}{l}{\textbf{$\chi^{2}(1)$}} & \multicolumn{1}{l}{\textbf{p}}        \\ \hline
\multicolumn{1}{|l|}{\multirow{6}{*}{Natural}}                                                                     & \multicolumn{1}{l|}{Gaze}                    & \multicolumn{1}{l|}{4 (Q1=3, Q3= 5)}       & \multicolumn{1}{l|}{4 (Q1=3.75, Q3= 4)}    & \multicolumn{1}{l|}{.4}             & \multicolumn{1}{l|}{\textgreater{}.05} \\ \cline{2-6} 
\multicolumn{1}{|l|}{}                                                                                             & \multicolumn{1}{l|}{\gazeF}         & \multicolumn{1}{l|}{4 (Q1=3.75, Q3= 5)}    & \multicolumn{1}{l|}{4 (Q1=3.75, Q3= 5)}    & \multicolumn{1}{l|}{0}              & \multicolumn{1}{l|}{\textgreater{}.05} \\ \cline{2-6} 
\multicolumn{1}{|l|}{}                                                                                             & \multicolumn{1}{l|}{GazeTouch}               & \multicolumn{1}{l|}{4 (Q1=3, Q3= 4)}       & \multicolumn{1}{l|}{4 (Q1=3, Q3= 4.25)}    & \multicolumn{1}{l|}{1}              & \multicolumn{1}{l|}{\textgreater{}.05} \\ \cline{2-6} 
\multicolumn{1}{|l|}{}                                                                                             & \multicolumn{1}{l|}{\gazetouchF}    & \multicolumn{1}{l|}{3 (Q1=3,Q3=4)}         & \multicolumn{1}{l|}{3 (Q1=3,Q3=4)}         & \multicolumn{1}{l|}{4.74}           & \multicolumn{1}{l|}{\textless{}.05*}   \\ \cline{2-6} 
\multicolumn{1}{|l|}{}                                                                                             & \multicolumn{1}{l|}{\touchOne}        & \multicolumn{1}{l|}{4 (Q1=3.75,Q3=5)}      & \multicolumn{1}{l|}{5 (Q1=4,Q3=5)}         & \multicolumn{1}{l|}{3.27}           & \multicolumn{1}{l|}{\textgreater{}.05} \\ \cline{2-6} 
\multicolumn{1}{|l|}{}                                                                                             & \multicolumn{1}{l|}{\touchTwo}        & \multicolumn{1}{l|}{4 (Q1=2,Q3=4)}         & \multicolumn{1}{l|}{4 (Q1=4,Q3=5)}         & \multicolumn{1}{l|}{15}             & \multicolumn{1}{l|}{\textless{}.0005*} \\ \hline
\multicolumn{1}{|l|}{\multirow{6}{*}{Fast}}                                                                        & \multicolumn{1}{l|}{Gaze}                    & \multicolumn{1}{l|}{4 (Q1=3, Q3= 5)}       & \multicolumn{1}{l|}{4 (Q1=2.75, Q3= 4)}    & \multicolumn{1}{l|}{1.33}           & \multicolumn{1}{l|}{\textgreater{}.05} \\ \cline{2-6} 
\multicolumn{1}{|l|}{}                                                                                             & \multicolumn{1}{l|}{\gazeF}         & \multicolumn{1}{l|}{4 (Q1=3, Q3= 5)}       & \multicolumn{1}{l|}{4 (Q1=4, Q3= 5)}       & \multicolumn{1}{l|}{.25}            & \multicolumn{1}{l|}{\textgreater{}.05} \\ \cline{2-6} 
\multicolumn{1}{|l|}{}                                                                                             & \multicolumn{1}{l|}{GazeTouch}               & \multicolumn{1}{l|}{3.5 (Q1=2, Q3= 4)}     & \multicolumn{1}{l|}{4 (Q1=3, Q3= 4.25)}    & \multicolumn{1}{l|}{.33}            & \multicolumn{1}{l|}{\textgreater{}.05} \\ \cline{2-6} 
\multicolumn{1}{|l|}{}                                                                                             & \multicolumn{1}{l|}{\gazetouchF}    & \multicolumn{1}{l|}{4 (Q1=3,Q3=4)}         & \multicolumn{1}{l|}{4 (Q1=3,Q3=4)}         & \multicolumn{1}{l|}{.33}            & \multicolumn{1}{l|}{\textgreater{}.05} \\ \cline{2-6} 
\multicolumn{1}{|l|}{}                                                                                             & \multicolumn{1}{l|}{\touchOne}        & \multicolumn{1}{l|}{4 (Q1=3,Q3=4.25)}      & \multicolumn{1}{l|}{5 (Q1=4,Q3=5)}         & \multicolumn{1}{l|}{5.4}            & \multicolumn{1}{l|}{\textless{}.05*}   \\ \cline{2-6} 
\multicolumn{1}{|l|}{}                                                                                             & \multicolumn{1}{l|}{\touchTwo}        & \multicolumn{1}{l|}{4 (Q1=3,Q3=4)}         & \multicolumn{1}{l|}{5 (Q1=4,Q3=5)}         & \multicolumn{1}{l|}{13.24}          & \multicolumn{1}{l|}{\textless{}.0005*} \\ \hline
\multicolumn{1}{|l|}{\multirow{6}{*}{Enjoyable}}                                                                   & \multicolumn{1}{l|}{Gaze}                    & \multicolumn{1}{l|}{4 (Q1=3, Q3= 5)}       & \multicolumn{1}{l|}{4 (Q1=3, Q3= 4.25)}    & \multicolumn{1}{l|}{.08}            & \multicolumn{1}{l|}{\textgreater{}.05} \\ \cline{2-6} 
\multicolumn{1}{|l|}{}                                                                                             & \multicolumn{1}{l|}{\gazeF}         & \multicolumn{1}{l|}{4 (Q1=3, Q3= 5)}       & \multicolumn{1}{l|}{4 (Q1=3.75, Q3= 5)}    & \multicolumn{1}{l|}{.33}            & \multicolumn{1}{l|}{\textgreater{}.05} \\ \cline{2-6} 
\multicolumn{1}{|l|}{}                                                                                             & \multicolumn{1}{l|}{GazeTouch}               & \multicolumn{1}{l|}{4 (Q1=3, Q3= 4)}       & \multicolumn{1}{l|}{4 (Q1=3.75, Q3= 4.25)} & \multicolumn{1}{l|}{2.27}           & \multicolumn{1}{l|}{\textgreater{}.05} \\ \cline{2-6} 
\multicolumn{1}{|l|}{}                                                                                             & \multicolumn{1}{l|}{\gazetouchF}    & \multicolumn{1}{l|}{3 (Q1=2, Q3= 4)}       & \multicolumn{1}{l|}{4 (Q1=3, Q3= 5)}       & \multicolumn{1}{l|}{3.6}            & \multicolumn{1}{l|}{\textgreater{}.05} \\ \cline{2-6} 
\multicolumn{1}{|l|}{}                                                                                             & \multicolumn{1}{l|}{\touchOne}        & \multicolumn{1}{l|}{3 (Q1=2.75,Q3=4)}      & \multicolumn{1}{l|}{4 (Q1=3,Q3=5)}         & \multicolumn{1}{l|}{8.05}           & \multicolumn{1}{l|}{\textless{}.005*}  \\ \cline{2-6} 
\multicolumn{1}{|l|}{}                                                                                             & \multicolumn{1}{l|}{\touchTwo}        & \multicolumn{1}{l|}{3 (Q1=2,Q3=3)}         & \multicolumn{1}{l|}{4 (Q1=3,Q3=4)}         & \multicolumn{1}{l|}{9.94}           & \multicolumn{1}{l|}{\textless{}.005*}  \\ \hline
\multicolumn{1}{|l|}{\multirow{6}{*}{Easy to use}}                                                                 & \multicolumn{1}{l|}{Gaze}                    & \multicolumn{1}{l|}{4 (Q1=3,Q3=4.25)}      & \multicolumn{1}{l|}{4 (Q1=3,Q3=4.25)}      & \multicolumn{1}{l|}{.82}            & \multicolumn{1}{l|}{\textgreater{}.05} \\ \cline{2-6} 
\multicolumn{1}{|l|}{}                                                                                             & \multicolumn{1}{l|}{\gazeF}         & \multicolumn{1}{l|}{4 (Q1=3.75,Q3=5)}      & \multicolumn{1}{l|}{4 (Q1=4,Q3=5)}         & \multicolumn{1}{l|}{1.14}           & \multicolumn{1}{l|}{\textgreater{}.05} \\ \cline{2-6} 
\multicolumn{1}{|l|}{}                                                                                             & \multicolumn{1}{l|}{GazeTouch}               & \multicolumn{1}{l|}{4 (Q1=2.75,Q3=4)}      & \multicolumn{1}{l|}{4 (Q1=3,Q3=5)}         & \multicolumn{1}{l|}{.33}            & \multicolumn{1}{l|}{\textgreater{}.05} \\ \cline{2-6} 
\multicolumn{1}{|l|}{}                                                                                             & \multicolumn{1}{l|}{\gazetouchF}    & \multicolumn{1}{l|}{4 (Q1=2,Q3=4)}         & \multicolumn{1}{l|}{4 (Q1=3,Q3=5)}         & \multicolumn{1}{l|}{4}              & \multicolumn{1}{l|}{\textless{}.05*}   \\ \cline{2-6} 
\multicolumn{1}{|l|}{}                                                                                             & \multicolumn{1}{l|}{\touchOne}        & \multicolumn{1}{l|}{4 (Q1=2,Q3=4)}         & \multicolumn{1}{l|}{5 (Q1=4,Q3=5)}         & \multicolumn{1}{l|}{9}              & \multicolumn{1}{l|}{\textless{}.005*}  \\ \cline{2-6} 
\multicolumn{1}{|l|}{}                                                                                             & \multicolumn{1}{l|}{\touchTwo}        & \multicolumn{1}{l|}{3 (Q1=2,Q3=4)}         & \multicolumn{1}{l|}{5 (Q1=4,Q3=5)}         & \multicolumn{1}{l|}{14.22}          & \multicolumn{1}{l|}{\textless{}.0005*} \\ \hline
\multicolumn{1}{|l|}{\multirow{6}{*}{Accurate}}                                                                    & \multicolumn{1}{l|}{Gaze}                    & \multicolumn{1}{l|}{3 (Q1=3,Q3=4)}         & \multicolumn{1}{l|}{3 (Q1=2,Q3=4)}         & \multicolumn{1}{l|}{.07}            & \multicolumn{1}{l|}{\textgreater{}.05} \\ \cline{2-6} 
\multicolumn{1}{|l|}{}                                                                                             & \multicolumn{1}{l|}{\gazeF}         & \multicolumn{1}{l|}{3 (Q1=2,Q3=4)}         & \multicolumn{1}{l|}{4 (Q1=3,Q3=5)}         & \multicolumn{1}{l|}{2.88}           & \multicolumn{1}{l|}{\textgreater{}.05} \\ \cline{2-6} 
\multicolumn{1}{|l|}{}                                                                                             & \multicolumn{1}{l|}{GazeTouch}               & \multicolumn{1}{l|}{3 (Q1=2.75,Q3=4)}      & \multicolumn{1}{l|}{4 (Q1=3,Q3=4)}         & \multicolumn{1}{l|}{5.44}           & \multicolumn{1}{l|}{\textless{}.05*}   \\ \cline{2-6} 
\multicolumn{1}{|l|}{}                                                                                             & \multicolumn{1}{l|}{\gazetouchF}    & \multicolumn{1}{l|}{3 (Q1=2,Q3=4)}         & \multicolumn{1}{l|}{4 (Q1=3,Q3=4)}         & \multicolumn{1}{l|}{5.4}            & \multicolumn{1}{l|}{\textless{}.05*}   \\ \cline{2-6} 
\multicolumn{1}{|l|}{}                                                                                             & \multicolumn{1}{l|}{\touchOne}        & \multicolumn{1}{l|}{5 (Q1=4,Q3=5)}         & \multicolumn{1}{l|}{5 (Q1=4.75,Q3=5)}      & \multicolumn{1}{l|}{3.6}            & \multicolumn{1}{l|}{\textgreater{}.05} \\ \cline{2-6} 
\multicolumn{1}{|l|}{}                                                                                             & \multicolumn{1}{l|}{\touchTwo}        & \multicolumn{1}{l|}{4 (Q1=4,Q3=5)}         & \multicolumn{1}{l|}{5 (Q1=5,Q3=5)}         & \multicolumn{1}{l|}{6.4}            & \multicolumn{1}{l|}{\textless{}.05*}   \\ \hline
\multicolumn{1}{|l|}{\multirow{6}{*}{\begin{tabular}[c]{@{}l@{}}Use for targets \\ in top region\end{tabular}}}    & \multicolumn{1}{l|}{Gaze}                    & \multicolumn{1}{l|}{4 (Q1=3.75,Q3=5)}      & \multicolumn{1}{l|}{4 (Q1=4,Q3=5)}         & \multicolumn{1}{l|}{.82}            & \multicolumn{1}{l|}{\textgreater{}.05} \\ \cline{2-6} 
\multicolumn{1}{|l|}{}                                                                                             & \multicolumn{1}{l|}{\gazeF}         & \multicolumn{1}{l|}{4 (Q1=4,Q3=5)}         & \multicolumn{1}{l|}{4 (Q1=4,Q3=5)}         & \multicolumn{1}{l|}{.82}            & \multicolumn{1}{l|}{\textgreater{}.05} \\ \cline{2-6} 
\multicolumn{1}{|l|}{}                                                                                             & \multicolumn{1}{l|}{GazeTouch}               & \multicolumn{1}{l|}{4 (Q1=3.75,Q3=4)}      & \multicolumn{1}{l|}{4 (Q1=4,Q3=5)}         & \multicolumn{1}{l|}{.6}             & \multicolumn{1}{l|}{\textgreater{}.05} \\ \cline{2-6} 
\multicolumn{1}{|l|}{}                                                                                             & \multicolumn{1}{l|}{\gazetouchF}    & \multicolumn{1}{l|}{4 (Q1=3,Q3=4)}         & \multicolumn{1}{l|}{4 (Q1=4,Q3=4)}         & \multicolumn{1}{l|}{1.67}           & \multicolumn{1}{l|}{\textgreater{}.05} \\ \cline{2-6} 
\multicolumn{1}{|l|}{}                                                                                             & \multicolumn{1}{l|}{\touchOne}        & \multicolumn{1}{l|}{3 (Q1=2,Q3=4)}         & \multicolumn{1}{l|}{4 (Q1=4,Q3=5)}         & \multicolumn{1}{l|}{5.56}           & \multicolumn{1}{l|}{\textless{}.05*}   \\ \cline{2-6} 
\multicolumn{1}{|l|}{}                                                                                             & \multicolumn{1}{l|}{\touchTwo}        & \multicolumn{1}{l|}{4 (Q1=3,Q3=4)}         & \multicolumn{1}{l|}{5 (Q1=4,Q3=5)}         & \multicolumn{1}{l|}{6.23}           & \multicolumn{1}{l|}{\textless{}.05*}   \\ \hline
\multicolumn{1}{|l|}{\multirow{6}{*}{\begin{tabular}[c]{@{}l@{}}Use for targets \\ in middle region\end{tabular}}} & \multicolumn{1}{l|}{Gaze}                    & \multicolumn{1}{l|}{4 (Q1=3,Q3=5)}         & \multicolumn{1}{l|}{4 (Q1=3,Q3=4)}         & \multicolumn{1}{l|}{1}              & \multicolumn{1}{l|}{\textgreater{}.05} \\ \cline{2-6} 
\multicolumn{1}{|l|}{}                                                                                             & \multicolumn{1}{l|}{\gazeF}         & \multicolumn{1}{l|}{4 (Q1=4,Q3=5)}         & \multicolumn{1}{l|}{4 (Q1=4,Q3=5)}         & \multicolumn{1}{l|}{.08}            & \multicolumn{1}{l|}{\textgreater{}.05} \\ \cline{2-6} 
\multicolumn{1}{|l|}{}                                                                                             & \multicolumn{1}{l|}{GazeTouch}               & \multicolumn{1}{l|}{4 (Q1=3,Q3=4.25)}      & \multicolumn{1}{l|}{4 (Q1=4,Q3=5)}         & \multicolumn{1}{l|}{3}              & \multicolumn{1}{l|}{\textgreater{}.05} \\ \cline{2-6} 
\multicolumn{1}{|l|}{}                                                                                             & \multicolumn{1}{l|}{\gazetouchF}    & \multicolumn{1}{l|}{4 (Q1=3,Q3=4.25)}      & \multicolumn{1}{l|}{4 (Q1=3,Q3=4)}         & \multicolumn{1}{l|}{.25}            & \multicolumn{1}{l|}{\textgreater{}.05} \\ \cline{2-6} 
\multicolumn{1}{|l|}{}                                                                                             & \multicolumn{1}{l|}{\touchOne}        & \multicolumn{1}{l|}{4 (Q1=4,Q3=5)}         & \multicolumn{1}{l|}{5 (Q1=4,Q3=5)}         & \multicolumn{1}{l|}{1.92}           & \multicolumn{1}{l|}{\textgreater{}.05} \\ \cline{2-6} 
\multicolumn{1}{|l|}{}                                                                                             & \multicolumn{1}{l|}{\touchTwo}        & \multicolumn{1}{l|}{4 (Q1=3.75,Q3=4)}      & \multicolumn{1}{l|}{4 (Q1=4,Q3=5)}         & \multicolumn{1}{l|}{5.33}           & \multicolumn{1}{l|}{\textless{}.05*}   \\ \hline
\multicolumn{1}{|l|}{\multirow{6}{*}{\begin{tabular}[c]{@{}l@{}}Use for targets \\ in bottom region\end{tabular}}} & \multicolumn{1}{l|}{Gaze}                    & \multicolumn{1}{l|}{4 (Q1=3,Q3=4)}         & \multicolumn{1}{l|}{4 (Q1=3,Q3=4)}         & \multicolumn{1}{l|}{.6}             & \multicolumn{1}{l|}{\textgreater{}.05} \\ \cline{2-6} 
\multicolumn{1}{|l|}{}                                                                                             & \multicolumn{1}{l|}{\gazeF}         & \multicolumn{1}{l|}{4 (Q1=3,Q3=4.25)}      & \multicolumn{1}{l|}{4 (Q1=4,Q3=4)}         & \multicolumn{1}{l|}{.33}            & \multicolumn{1}{l|}{\textgreater{}.05} \\ \cline{2-6} 
\multicolumn{1}{|l|}{}                                                                                             & \multicolumn{1}{l|}{GazeTouch}               & \multicolumn{1}{l|}{4 (Q1=2.75,Q3=4)}      & \multicolumn{1}{l|}{4 (Q1=3,Q3=4)}         & \multicolumn{1}{l|}{0}              & \multicolumn{1}{l|}{\textgreater{}.05} \\ \cline{2-6} 
\multicolumn{1}{|l|}{}                                                                                             & \multicolumn{1}{l|}{\gazetouchF}    & \multicolumn{1}{l|}{3.5 (Q1=2.75,Q3=4)}    & \multicolumn{1}{l|}{4 (Q1=3,Q3=4)}         & \multicolumn{1}{l|}{.47}            & \multicolumn{1}{l|}{\textgreater{}.05} \\ \cline{2-6} 
\multicolumn{1}{|l|}{}                                                                                             & \multicolumn{1}{l|}{\touchOne}        & \multicolumn{1}{l|}{3.5 (Q1=2.75,Q3=4.25)} & \multicolumn{1}{l|}{4 (Q1=4,Q3=5)}         & \multicolumn{1}{l|}{3.27}           & \multicolumn{1}{l|}{\textgreater{}.05} \\ \cline{2-6} 
\multicolumn{1}{|l|}{}                                                                                             & \multicolumn{1}{l|}{\touchTwo}        & \multicolumn{1}{l|}{4 (Q1=3,Q3=4)}         & \multicolumn{1}{l|}{4 (Q1=4,Q3=5)}         & \multicolumn{1}{l|}{11.27}          & \multicolumn{1}{l|}{\textless{}.001*}  \\ \hline
\multicolumn{1}{|l|}{\multirow{6}{*}{Overall use}}                                                                 & \multicolumn{1}{l|}{Gaze}                    & \multicolumn{1}{l|}{4 (Q1=3,Q3=5)}         & \multicolumn{1}{l|}{3 (Q1=3,Q3=4)}         & \multicolumn{1}{l|}{1}              & \multicolumn{1}{l|}{\textgreater{}.05} \\ \cline{2-6} 
\multicolumn{1}{|l|}{}                                                                                             & \multicolumn{1}{l|}{\gazeF}         & \multicolumn{1}{l|}{4 (Q1=3,Q3=5)}         & \multicolumn{1}{l|}{4 (Q1=3,Q3=4)}         & \multicolumn{1}{l|}{1.14}           & \multicolumn{1}{l|}{\textgreater{}.05} \\ \cline{2-6} 
\multicolumn{1}{|l|}{}                                                                                             & \multicolumn{1}{l|}{GazeTouch}               & \multicolumn{1}{l|}{4 (Q1=3,Q3=4)}         & \multicolumn{1}{l|}{4 (Q1=3.75,Q3=4.25)}   & \multicolumn{1}{l|}{1.32}           & \multicolumn{1}{l|}{\textgreater{}.05} \\ \cline{2-6} 
\multicolumn{1}{|l|}{}                                                                                             & \multicolumn{1}{l|}{\gazetouchF}    & \multicolumn{1}{l|}{4 (Q1=3,Q3=4)}         & \multicolumn{1}{l|}{4 (Q1=3,Q3=4)}         & \multicolumn{1}{l|}{.69}            & \multicolumn{1}{l|}{\textgreater{}.05} \\ \cline{2-6} 
\multicolumn{1}{|l|}{}                                                                                             & \multicolumn{1}{l|}{\touchOne}        & \multicolumn{1}{l|}{4 (Q1=3,Q3=4)}         & \multicolumn{1}{l|}{4 (Q1=4,Q3=5)}         & \multicolumn{1}{l|}{4.76}           & \multicolumn{1}{l|}{\textless{}.05*}   \\ \cline{2-6} 
\multicolumn{1}{|l|}{}                                                                                             & \multicolumn{1}{l|}{\touchTwo}        & \multicolumn{1}{l|}{4 (Q1=3,Q3=5)}         & \multicolumn{1}{l|}{4 (Q1=4,Q3=5)}         & \multicolumn{1}{l|}{3.27}           & \multicolumn{1}{l|}{\textgreater{}.05} \\ \hline
\end{tabular}
\end{table*}

\end{document}